\documentclass[a4paper]{article}
\usepackage{jcappub}
\usepackage{tabularx}
\usepackage{bibentry}    
\usepackage{amsfonts}
\usepackage{amsmath}
\usepackage{amssymb}     
\usepackage{mathrsfs}    
\usepackage{mathtools}
\usepackage{dsfont}      
\usepackage{xcolor}      
\usepackage{xfrac}       
\usepackage{fontawesome}
\usepackage{cleveref}
\usepackage{listings}
\usepackage{comment}
\usepackage{cancel}
\usepackage{xspace}

\pdfoutput=1

\definecolor{shadecolor}{rgb}{0.95,0.95,0.95}
\newcommand{\Tkd}{T_\mathrm{kd}}

\newcommand{\eqsp}{\;\,}

\newcommand{\dm}{\delta_R}

\newcommand{\acropolis}{\texttt{ACROPOLIS}\xspace}

\title{Big Bang Nucleosynthesis constraints on resonant DM annihilations}
\author[a,b]{Pieter Braat}
\author[c]{and Marco Hufnagel}
\date{January 2023}
\emailAdd{pbraat@nikhef.nl}
\emailAdd{mail@hep-mh.com}
\affiliation[a]{Nikhef, Theory Group, Science Park 105, 1098 XG Amsterdam, The Netherlands}
\affiliation[b]{Institute of Physics, University of Amsterdam, Science Park 904, 1098 XH Amsterdam, The Netherlands}
\affiliation[c]{Service de Physique Théorique, Université Libre de Bruxelles, Boulevard du Triomphe, CP225, 1050 Brussels, Belgium}

\abstract{
We perform a systematic study of BBN constraints from photodisintegration for scenarios in which dark-matter annihilations are resonantly-enhanced. To this end, we implement and make available a new class \texttt{ResonanceModel} within an updated version v1.3.0 of \acropolis. While the corresponding implementation is done in a rather model-independent way, we also make available three benchmark models that can be used to calculate constraints for more concrete scenarios. Using this new version of \acropolis, we present for the first time the corresponding constraints on resonantly-enhanced $s$-wave and $p$-wave annihilations. We show that for $s$-wave annihilations the bounds are usually very similar to the ones without a resonance, while for $p$-wave annihilations the bounds can be significantly stronger.\newline\newline
The updated version v1.3.0 of \acropolis can be found at \url{https://github.com/hep-mh/acropolis}.
}

\begin{document}

\preprint{Nikhef-2023-014, ULB-TH/24-08}

\maketitle

\section{Introduction}
\label{sec:intro}
Despite the overwhelming evidence for its existence, a conclusive particle description of dark matter (DM) has yet to be found. In fact, the standard, minimal WIMP paradigm is increasingly constrained by direct-detection experiments~\cite{LZ:2022lsv,XENON:2023cxc,PandaX:2022osq} and as a result more exotic dark sectors (DS) are currently being explored. In particular, setups in which DM annihilations are resonantly enhanced by additional DS states have recently gained a lot of attention~\cite{Bernreuther:2020koj,Tsai:2020vpi,Csaki:2022xmu,Kondo:2022lgg,Binder:2022pmf,Beneke:2022rjv,Hryczuk:2022gay,Kim:2022cpu,Brahma:2023psr,Balan:2024cmq,Braat:2023fhn,Garcia-Cely:2024ivo}. If a DS contains a mass state $R$ that has approximately twice the mass of the DM particle $\chi$, i.e.
\begin{equation}
    m_R = m_\chi (2+\dm) \qquad \text{with} \qquad \dm  \ll 1\eqsp,
    \label{eq:def_eps}
\end{equation}
DM annihilations into Standard Model (SM) states of the form $\chi\bar\chi \rightarrow R \rightarrow \text{SM}\,\text{SM}$ are resonantly enhanced. If this enhancement happens during the time of DM freeze-out, it is possible to obtain the correct relic abundance for comparatively small couplings~\cite{Bernreuther:2020koj,Braat:2023fhn,Brahma:2023psr}, thus avoiding some of the experimental constraints that plague the usual WIMP scenario.
In addition, small-scale structure problems, such as the cusp vs.~core problem~\cite{Kaplinghat:2015aga} and the diversity problem~\cite{Oman:2015xda,Ren:2018jpt} suggest that DM self-interactions be velocity dependent, in order to solve the discrepancy between the observed strengths of self-interactions at the scales of galaxies and galaxy clusters~\cite{Chu:2019awd,Correa:2020qam}. Initially, this velocity-dependence was accommodated for via the inclusion of light dark mediators~\cite{Tulin:2013teo,Hambye:2019tjt}. However, it has recently been shown that a similar effect can also be achieved if the DM self-interaction cross-section is \textit{velocity independent} but instead \textit{resonantly enhanced} for typical velocities in galaxy-sized DM halos~\cite{Chu:2018fzy,Braat:2023fhn,Garcia-Cely:2024ivo}. To accommodate this effect, rather small mass splitting are required, i.e.~$\dm\sim10^{-8}$ for $s$- and/or $p$-wave resonances.

One implication of such scenarios is that residual DM annihilations -- i.e.~those annihilations that proceed out-of-equilibrium even after DM freeze-out has already concluded -- can still have a significant impact at late times due to the increasingly efficient annihilation process~\cite{Braat:2023fhn}. In the early Universe, these annihilations might for example still inject a significant amount of electromagnetic energy into the SM heat bath, which can be constrained by cosmological observations including those from Big Bang Nucleosynthesis (BBN)~\cite{Pospelov:2010hj,Hufnagel:2018bjp,Forestell:2018txr,Depta:2019lbe,Depta:2020mhj,Depta:2020zbh,Coffey:2020oir,Alves:2023jlo} and the Cosmic Microwave Background (CMB)~\cite{Slatyer:2015jla,Slatyer:2015kla}. For non-resonant annihilations, the resulting CMB bounds are typically found to be dominant for $s$-wave annihilations~\cite{Bringmann:2021sth}, whereas the corresponding BBN observations outperform the CMB ones in the case of $p$-wave annihilations~\cite{Depta:2019lbe}. The latter constraints have previously already been studied for GeV-scale~\cite{Hisano:2009rc,Jedamzik:2009uy,Kawasaki:2015yya} and MeV-scale particles~\cite{Depta:2019lbe}. 
However, in these works the annihilation cross-section was assumed to be non-resonant, i.e.~of the form $\langle \sigma_\text{ann} v \rangle = a + b\langle v^2 \rangle$ with some constants $a$ and $b$, describing $s$- and $p$-wave annihilations, respectively. Additionally, velocity dependent cross sections of the form $\langle \sigma_\text{ann} v \rangle \propto v^{-n}$ have been studied for fixed $n=1$ and $n=2$, which describe Sommerfeld and -- to some extent -- Breit-Wigner enhanced annihilations \cite{Hisano:2011dc}. The explicit connection to dark sectors with a resonant state has not been made, since the corresponding annihilation cross section cannot be described by a fixed $n$-scaling.
Thus, either form of the cross-section is not applicable to the case of resonant annihilations, meaning that the resulting constraints are not yet known. Calculating these limits will be the main focus of this work.

In a nutshell, the term BBN describes a process in the early Universe during which the first light nuclei were synthesized. Remarkably, when modeling this process within the $\Lambda$CDM framework, the predicted abundances of the various light elements agree exceptionally well with cosmological observations~\cite{Workman:2022ynf}. On the flip-side, any model featuring processes that go beyond the $\Lambda$CDM paradigm therefore have the potential to spoil this agreement, which can be used to put constraints on non-$\Lambda$CDM model parameters. For the scenarios considered in this work, there are two main effects that can influence the formation of light elements: On the one hand, any particle in the DS that is still relativistic at the time of BBN significantly changes the expansion history of the Universe and thus the final composition of light elements. This leads to a model-independent bound on the presence of light relics, which excludes DM masses below $\sim10\,\mathrm{MeV}$ and has already been studied in the literature~\cite{Depta:2019lbe}. We will therefore not reproduce this bound in the present work. On the other hand, (resonantly-enhanced) residual DM annihilations might inject large amounts of electromagnetic material into the SM plasma after BBN has already concluded. If these non-thermal annihilation products carry enough energy, they can efficiently break apart the previously created elements via the process of photodisintegration (cf.~\cite{Kawasaki_1995,Poulin:2015opa,Forestell:2018txr} and references therein), thus leading to abundances that are potentially in conflict with observation. Calculating the resulting constraints from photodisintegration is a main focus of this work.

There already exists the public code \acropolis~\cite{Depta:2020zbh,Depta:2020mhj, Hufnagel:2018bjp}, which can be used to derive photodisintegration constraints for a variety of different scenarios. However, when it comes to residual DM annihilations, as of v1.2.2, \acropolis can only handle non-resonant annihilations. With this work, we therefore update \acropolis to v1.3.0, which is made publicly available and -- besides other improvements -- includes a model that allows to calculate photodisintegration constraints for resonantly-enhanced DM annihilations in a programmatic manner. The implementation is thereby done in a rather model-independent way, by only making minor assumptions about the DS. Using this model, we then calculate the corresponding constraints for $s$- and $p$-wave annihilations, and show that especially in the latter case, resonantly-enhanced annihilations can lead to stronger constraints than in the vanilla, i.e.~non-resonant, scenario.

This paper is structured as follows: In sec.~\ref{sec:photodisintegration}, we review the most important steps that are necessary to calculate constraints from photodisintegration as implemented in \acropolis. In sec.~\ref{sec:model}, we present the model-independent setup that we use to model resonant DM annihilations, and in sec.~\ref{sec:implementation}, we discuss our implementation of this setup within \acropolis. In sec.~\ref{sec:results}, we present example results obtained via this implementation for $s$- and $p$-wave annihilations. Finally, we conclude in sec.~\ref{sec:conclusion}. The updated version of \acropolis is made available at \url{https://github.com/hep-mh/acropolis}.

\section{The basic principles of photodisintegration}
\label{sec:photodisintegration}
In this section, we first review the basic principles of photodisintegration as well as the steps that are needed to calculate the corresponding constraints. Thereby, we put special emphasis on residual DM annihilations as well as the relevant formulae that are used within \acropolis (for more details, see~\cite{Depta:2020mhj}). 

In the early Universe, any electromagnetic material that is injected into the SM heat bath via processes of the form $X \rightarrow \gamma\gamma$ or $X \rightarrow e^+ e^-$ will induce an electromagnetic cascade via interactions with the background plasma. More precisely, denoting thermal background particles with a subscript $\cdot_\text{th}$, injected photons will scatter with the background via electron-positron pair creation $\gamma \gamma_{\mathrm{th}}\to e^+e^-$, photon-photon scattering $\gamma \gamma_{\mathrm{th}}\to \gamma\gamma$, pair creation on nuclei $\gamma N \to N e^+ e^-$ and/or Compton scattering $\gamma e^-_{\mathrm{th}} \to \gamma e^-$, while injected electrons/positrons mainly undergo inverse Compton scattering $e^\pm \gamma_{\mathrm{th}} \to e^\pm \gamma$. As a result of this cascade, the injected particles $X \in \{e^-,e^+,\gamma \}$\footnote{Sufficiently heavy DM particles can also inject hadronic particles like pions and nucleons, which can additionally influence the neutron-to-proton ratio and participate in hadrodisintegration reactions. These effects have mostly been studied in the case of DM decays into hadrons \cite{Reno:1987qw,Kawasaki:2004qu,Jedamzik:2006xz,Freitas:2009jb,Kawasaki:2017bqm}, in which case the bounds are stronger compared to exclusive EM final states \cite{Boyarsky:2020dzc}. In the case of annihilations, such effects become relevant for $m_\chi > m_\pi$ and we expect them to lead to even stronger constraints \cite{Hisano:2009rc}. However, in this work, we assume that the DM annihilations exclusively inject electromagnetic material, and we leave the study of the effects of potential hadronic injections for future work.} assume characteristic non-thermal spectra $\mathrm{f}_X$,\footnote{These spectra are normalized in such a way that $\int \mathrm{f}_X \text{d} E = n_X$.} which can be determined by solving the cascade equation (dropping the $T$ dependence for convenience)~\cite{Hufnagel:2018bjp,Jedamzik:2006xz,Kawasaki_1995}
\begin{equation}
    \mathrm{f}_X(E) = \frac{1}{\Gamma_X(E)} \left( S_X(E) + \int_E^\infty dE' \sum_{X'} K_{X'\to X}(E,E') \mathrm{f}_{X'}(E') \right)\eqsp.
    \label{eq:cascade}
\end{equation}
Here, $\Gamma_X(E)$ is the total scattering rate of particle $X$ at energy $E$, $K_{X'\rightarrow X}(E, E')$ is the differential scattering rate of particle $X'$ with energy $E'$ into particle $X$ with energy $E$, and $S_X(E)$ is the source term describing the amount of injected $X$ particles with energy $E$.

Notably, the source term $S_X(E)$ depends on the exact type of the injection and is usually split into a monochromatic part $S^{(0)}$ and a continuous part $S^{(c)}$, i.e.
\begin{equation}
    S_X(E) = S_X^{(0)} \delta (E-E_0) + S_{X}^{(c)}(E)\eqsp,
\end{equation}
with the maximal injection energy $E_0$.
In the case of DM annihilations, i.e.~for injections of the form $\chi\bar\chi\rightarrow \gamma \gamma$ or $\chi\bar\chi \rightarrow e^+ e^-$, we have $E_0 = m_\chi$.\footnote{Since residual annihilations happen after DM freeze-out, the DM particles can be assumed to be at rest.} In this case, the monochromatic source terms are given by the product of the thermally averaged annihilation cross-section $\langle \sigma_\text{ann} v \rangle$ and the square of the DM number density $n_\chi$~\cite{Kawasaki_1995, Depta:2019lbe}, i.e.
\begin{align}
    S_{e^-}^{(0)} = S_{e^+}^{(0)} = \frac12 n_\chi^2 \langle \sigma_\text{ann} v \rangle_{\chi \chi \to e^+e^-}\quad \text{and} \qquad S_{\gamma}^{(0)} = n_\chi^2 \langle \sigma_\text{ann} v \rangle_{\chi \chi \to \gamma\gamma}\eqsp.
    \label{eq:source_terms}
\end{align}
Additionally, the continuous source terms are non-zero only for photons and encode the additional energy injection due to final-state radiation of the form $\chi \bar\chi \to e^+e^-\gamma$~\cite{Forestell:2018txr},
\begin{align}
    S_{\gamma}^{(c)}(E) = \frac{n_\chi^2 \langle \sigma_\text{ann} v \rangle_{\chi \chi \to e^+e^-}}{2m_\chi} \frac{\alpha}{\pi} \frac{1+(1-y)^2}{y}\ln\left(\frac{4m_\chi^2(1-y)}{m_e^2} \right) \Theta\left(1-\frac{m_e^2}{4m_\chi^2}-y \right)\eqsp,
\end{align}
with $y=E/m_\chi$. By using the given source terms as an input, the non-thermal photon spectrum resulting from the electromagnetic cascade is then obtained as the solution of eq.~\eqref{eq:cascade}. After the cascade, the non-thermal photons will engage in photodisintegration reactions of the form $\gamma N \rightarrow [\dots]$ with any of the light elements $N\in \{n, p, \mathrm{D, {}^3H, {}^3He, {}^4He, {}^7Li, {}^7Be}\}$. The resulting evolution of the different abundances $Y_N\equiv n_N/n_b$ normalized to the baryon-density $n_b$, is then obtained by solving the Boltzmann equation
~\cite{Depta:2019lbe,Poulin:2015opa,Cyburt:2002uv}
\begin{align}
    \frac{\text{d}T}{\text{d}t} \frac{\text{d}Y_X}{\text{d} T} = \sum_{N_i} Y_{N_i} \int_0^\infty \text{d}E\, \mathrm{f}_\gamma (E) \sigma_{\gamma+N_i \to X}(E) - Y_X \sum_{N_f}\int_0^\infty \text{d}E\, \mathrm{f}_\gamma(E) \sigma_{\gamma+X \to N_f}(E)\eqsp,
    \label{eq:Yx}
\end{align}
with $\sigma_{\gamma+[\dots] \to [\dots]}(E)$ being the cross-sections for the various disintegration reactions (for a list of all relevant reactions, see e.g.~\cite{Depta:2020mhj}).
Given a set of initial abundances $Y_X^{(0)}$ -- in this work, we use the abundances resulting from standard BBN --, as well as the spectrum $\mathrm{f}_\gamma(E)$ from eq.~\eqref{eq:cascade}, the solution of this equation then predicts the final abundances of the various elements after the process of photodisintegration, which can afterwards be compared to observations. Here, we employ the most recent measurements as implemented in \acropolis, i.e.
\begin{align}
    &\mathcal{Y}_p \qquad &0.245 \pm 0.003\eqsp, \quad \text{\cite{Workman:2022ynf}} \\
    &\mathrm{D}/^1\mathrm{H} \qquad &(25.47 \pm 0.25)\times 10^{-6}\eqsp, \quad \text{\cite{Workman:2022ynf}} \\
    &^3\mathrm{He}/\mathrm{D} \qquad &(8.3\pm1.5)\times 10^{-1}\eqsp. \quad \text{\cite{geiss:2003iso}}
\end{align}

Given an implementation of the relevant source terms in eq.~\eqref{eq:source_terms}, which depend on the underlying injection mechanism, all of the above steps can be handled by \acropolis. Specifically, \acropolis can solve both eqs.~\eqref{eq:cascade} and~\eqref{eq:Yx}, and afterwards compare the resulting abundances with the most recent observations. In the process, the code can also incorporate theoretical uncertainties on the nuclear reactions rates by running the same calculation with three different sets of initial conditions, which have been calculated using the mean, low and high values of the rates in question. The theoretical and experimental uncertainties can then be combined to infer the resulting limits at $95\%$ C.L. Regarding the source terms, however, as of v1.2.2, \acropolis only provides implementations for residual DM annihilations with cross-sections of the form $\langle \sigma_\text{ann} v \rangle = a + b\langle v^2 \rangle$ and constants $a$, $b$, which is not a valid parametrization in case of resonantly-enhanced annihilations. Consequently, in order to fully make use of the given machinery, we have to replace the annihilation cross-section implemented in \acropolis with one that is more suitable for resonantly-enhanced annihilations.

\section{Model description}
\label{sec:model}

\subsection{The annihilation cross-section}
In this paper, we employ a fairly model independent description of resonant DM annihilations; however, for concreteness, we will later also consider three specific benchmark scenarios (cf.~tab.~\ref{tab:models}). Following~\cite{Chu:2018fzy}, the total cross-section $\sigma_\text{ann}$ for resonant annihilations of non-relativistic (NR) DM particles can be written in the Breit-Wigner form~\cite{Breit:1936}
\begin{align}
     \sigma_{\text{ann}}(v) =
           \frac{4\pi S}{m_\chi E(v)}
           \frac{ \Gamma_d(v) \Gamma_v(v_f)/4}{  [E(v)-E(v_R)]^2 + {\Gamma(v)^2}/{4}}\eqsp.
           \label{eq:sigma_fv}
\end{align}
Here, $v$ is the relative velocity between the initial-state DM particles, $v_f = v_f(v)$ is the \textit{relative} velocity between the final-state SM particles, and $v_\text{R} = 2\sqrt{\dm}$. Moreover, $\Gamma_v(v)$ and $\Gamma_d(v)$ are the partial (running) decay widths into visible- and dark-sector states, respectively, $\Gamma(v) = \Gamma_v(v) + \Gamma_d(v)$, and $S$ is a symmetry factor. Using $E(v) = m_\chi v^2 / 4$ as well as the \textit{individual} DM momentum $p = m_\chi v/2$, eq.~\eqref{eq:sigma_fv} takes the alternative form
\begin{align}
    \sigma_\text{ann}(p) = \frac{4\pi S}{p^2} \frac{m_\chi^2 \Gamma_d(p) \Gamma_v(p_f)/4}{(p^2-p_R^2)^2 + m_\chi^2\Gamma(p)^2/4}
    \label{eq:sigma_fp}
\end{align}
with $p_\text{R} = m_\chi\sqrt{\dm}$, $p_f(p) = \sqrt{m_\chi^2+p^2-m_f^2}$, and the final-state particle mass $m_f$. Adapting the parametrization from \cite{Chu:2018fzy}, the (partial) decay widths can be written as
\begin{align}
    \Gamma_v(p) = \gamma_v m_R \left(\frac{p}{m_\chi}\right)^{2n_v +1}
    \qquad\text{and}\qquad
    \Gamma_d(p) = \gamma_d m_R \left(\frac{p}{m_\chi}\right)^{2n_d +1}\eqsp.
    \label{eq:gammas}
\end{align}
Here, $\gamma_v$ and $\gamma_d$ are some coupling constants that depend on the underlying model parameters, and the parameters $n_v$ and $n_d$ distinguish between $s$-wave ($n_\cdot = 0)$ and $p$-wave annihilations ($n_\cdot = 1$). In the following, we will for simplicity assume that the DM particles annihilate predominately into electrons, in which case $n_v = 1$.

Given the cross-section from eq.~\eqref{eq:sigma_fp} and assuming Maxwell-Boltzmann statistics for all interacting particles, the corresponding thermally averaged cross-section is given by~\cite{Gondolo:1990dk}
\begin{align}
    \langle \sigma_\text{ann} v \rangle = \frac{x}{8m_\chi^5 K_2^2(x)} \int_{4m_\chi^2}^\infty \mathrm{d}s\;  \sigma_\text{ann}(s) (s-4m_\chi^2)\sqrt{s} K_1(\sqrt{s}x/m_\chi)
    \label{eq:thavg_1}
\end{align}
with the DM temperature $T_\chi = m_\chi/x$ and the center-of-mass energy $s/4 = m_\chi^2 + p^2$. For non-relativistic DM particles, we have $x \gg 1$ and $p \ll m_\chi$, in which case $K_1(x) \simeq K_2(x) \simeq \sqrt{\pi/(2x)}e^{-x}$ as well as $\sqrt{s} \simeq 2m_\chi + p^2/m_\chi$. Consequently, eq.~\eqref{eq:thavg_1} simplifies to
\begin{align}
    \langle \sigma_\text{ann} v \rangle \simeq \frac{4x^{3/2}}{m_\chi^4 \sqrt{\pi}}\int_0^\infty \mathrm{d}p^2\; p^2 \sigma_\text{ann}(p) e^{-p^2 x/m_\chi^2}\eqsp.
    \label{eq:sigv_general}
\end{align}
This expression directly maps onto the source terms in eq.~\eqref{eq:source_terms}, and given that we focus on DM annihilations into electron-positron pairs, we can identify $\langle \sigma_\text{ann} v \rangle_{\chi \chi \to \gamma\gamma} = 0$ and $\langle \sigma_\text{ann} v \rangle_{\chi \chi \to e^+e^-} = \langle \sigma_\text{ann} v\rangle$.

While eq.~\eqref{eq:sigv_general} can be used to calculate the thermally averaged annihilation cross-section in a general manner, it can also be further simplified within certain limits. In the narrow-width approximation (NWA), which is valid for $\Gamma(p_R)/m_R \ll 1$, eq.~\eqref{eq:sigma_fp} becomes
\begin{align}
    \sigma_\text{ann}^\mathrm{res}(p) \overset{\mathrm{NWA}}{\simeq} \frac{8\pi^2 S}{p^2} \frac{m_\chi \Gamma_d(p) \Gamma_v(p_f)/4}{\Gamma(p)} \delta(p^2-p_R^2)\eqsp,
    \label{eq:sig_ann}
\end{align}
which implies that the thermally averaged annihilation cross-section \textit{around the resonance} is given by
\begin{align}
    \langle \sigma_\text{ann} v \rangle^{\mathrm{res}} &\simeq \frac{8S (\pi x)^{3/2}}{m_\chi^3} \frac{\Gamma_d(p) \Gamma_v(p_f)}{\Gamma(p)}\Big|_{p = p_R} e^{-\dm x} \nonumber \\
    &= \frac{8S (\pi x)^{3/2} m_R^2}{m_\chi^3} \frac{\gamma_v \gamma_d \dm^{n_d+1/2}}{\Gamma(p_R)} e^{-\dm x}\eqsp.
    \label{eq:sigma_res}
\end{align}
In the last step, we have used $p_f(p_R) = \sqrt{m_\chi^2(1+\dm) + m_f^2} \simeq m_\chi$ for $m_\chi/m_f \gg 1$ and consequently $\Gamma_v(p_f(p_R)) \simeq \gamma_v m_R$ as well as $\Gamma_d(p_R) = \gamma_d m_R \dm^{n_d + 1/2}$. \textit{Far away from the resonance}, i.e.~once the DM velocity has dropped significantly below the resonance velocity, we instead have $(p^2 - p_R^2)^2 \simeq p_R^4 = m_\chi^4 \dm^2 \gg \Gamma(p)$, meaning that the non-resonance contribution to the cross-section is approximately given by
\begin{align}
    \sigma_\text{ann}^\text{non-res}(p) \simeq \frac{\pi S}{p^2} \frac{\Gamma_d(p)\Gamma_v(p_f)}{m_\chi^2 \dm^2}\eqsp.
\end{align}
Plugging this expression back into eq.~\eqref{eq:sigv_general} and using $p_f(p) \simeq m_\chi$, we then find
\begin{align}
    \langle \sigma_\text{ann} v \rangle^{\mathrm{non-res}} \simeq \frac{4S\sqrt{\pi}\gamma_v \gamma_d m_R^2}{m_\chi^4\dm^2}x^{-n_d} \bar\Gamma(n_d+3/2)\eqsp,
    \label{eq:sigma_non_res}
\end{align}
with $\bar\Gamma$ being the gamma function. As expected, we obtain $\langle \sigma_\text{ann} v \rangle^{\mathrm{non-res}} \propto x^0 = \text{const}$ in the case of $s$-wave annihilations with $n_d = 0$, as well as $\langle \sigma_\text{ann} v \rangle^{\mathrm{non-res}} \propto x^{-1} = T_\chi/m \propto \langle v^2 \rangle$ in the case of $p$-wave annihilations with $n_d = 1$. More precisely,
\begin{align}
    n_d = 0: \qquad \langle \sigma_\text{ann} v \rangle^{\mathrm{non-res}} &= \frac{2S\pi\gamma_v \gamma_d}{m_\chi^2} \frac{(2+\dm)^2}{\dm^2}\hphantom{\frac{1}{x}} \equiv a \label{eq:defa}\\
    n_d = 1: \qquad \langle \sigma_\text{ann} v \rangle^{\mathrm{non-res}} &= \frac{3S\pi\gamma_v \gamma_d}{m_\chi^2} \frac{(2+\dm)^2}{\dm^2} \frac{1}{x} \equiv \frac{6b}{x}\eqsp.\label{eq:defb}
\end{align}
Using the latter two equations, it is possible to map the parameters $\gamma_v$, $\gamma_d$, $\dm$, and $S$ onto the constants $a$ and $b$, which are usually used to parameterize non-resonant DM annihilations via $\langle \sigma_\text{ann} v \rangle = a + b\langle v^2 \rangle$. We will make use of this relation again at a later point (cf.~sec.~\ref{sec:results}) in order to establish a meaningful comparison between resonant and non-resonant DM annihilations.

At this point, it is worth noting that our description so far has been rather model-independent. To utilize the above formulae for a concrete scenario, it is thus necessary to deduce the values of $n_d$, $S$, $\gamma_d$, and $\gamma_v$ from the actual model parameters. For this work, we have performed this deduction for three different benchmark models, namely for \textit{(1)} a scalar DM particle $\varphi$ with a scalar resonance $\Phi$, \textit{(2)} a fermionic DM particle $\psi$ with a vector resonance $A'_\mu$, and \textit{(3)} a (complex) scalar DM particle $\varphi$ with a vector resonance $A'_\mu$. The corresponding parameter relations for these benchmark models are summarised in tab.~\ref{tab:models}.

\begin{table}[t]
    \centering
    \begin{tabular}{c|c|c|c|c|c|c|c|c}
        ID & model & $\chi$ & $R$ & Lagrangian & $n_d$ & $\gamma_d$ & $\gamma_v$ & $S$ \\ \hline
        (1) & 2 Scalars ($\varphi$ + $\Phi$) & $\varphi$ & $\Phi$ & $g_1 \varphi\varphi\Phi +g_2 \bar e e \Phi$ & 0 & $\frac{g_1^2}{64\pi m^2_\varphi}$ & $\frac{g_2^2}{8\pi}$ & $\frac12$ \\
        (2) & Fermion ($\psi$) + Vector ($A'_\mu$) & $\psi$ & $A'_\mu$ & $g_1 \bar \psi \gamma^\mu \psi A'_\mu + g_2 \bar e \gamma^\mu e A'_\mu$ & 0 & $\frac{g_1^2}{8\pi}$ & $\frac{g_2^2}{12\pi}$ & $\frac34$ \\
        (3) & Scalar ($\varphi$) + Vector ($A'_\mu$) & $\varphi$ & $A'_\mu$ & $g_1\varphi^\dagger {\overset{\leftrightarrow}{\partial}}_{\!\mu} \varphi A'^\mu + g_2 \bar e \gamma^\mu e A'_\mu$ & 1 &$\frac{g_1^2}{48\pi}$ & $\frac{g_2^2}{12\pi}$ & $\frac32$
    \end{tabular}
    \caption{Overview of the three benchmark models with UV parameters linked to the dimensionless parameters $\gamma_d$ and $\gamma_v$. Here, we explicitly assume that the particle $R$ couples exclusively to electrons.}
    \label{tab:models}
\end{table} 


\subsection{The dark-sector temperature}
\label{sec:ds_temp}
To evaluate the thermally averaged annihilation cross-section in eq.~\eqref{eq:sigv_general}, we still need to know the evolution of the DS temperature $T_\chi$, which critically depends on the temperature $\Tkd$ at which the DM particles decouple kinetically from the SM heat bath, i.e.~\cite{Depta:2019lbe,Kolb:1990vq}
\begin{equation}
    T_\chi(T) = \begin{cases}
        T & \mathrm{for}\quad T \geq \Tkd \\
        \Tkd R(\Tkd)^2/R(T)^2 & \mathrm{for}\quad T < \Tkd
    \end{cases}\eqsp,
\end{equation}
with the scale factor $R$ and $R(\Tkd)^2/R(T)^2 = T^2/\Tkd^2$ since photodisintegration is only relevant for $T < \mathcal{O}(1)\,\mathrm{keV} \ll m_e$~\cite{Hufnagel:2018bjp}.

In order to calculate $\Tkd$, let us recall that we focus on scenarios with DM annihilations that proceed exclusively into electron-positron pairs (cf.~tab.~\ref{tab:models}), in which case kinetic equilibrium between $\chi$ and the SM heat bath is maintained via reactions of the form $\chi e^\pm \leftrightarrow \chi e^\pm$. Following \cite{Bringmann:2006mu}, we \textit{approximate} $\Tkd$ by comparing the Hubble rate $H$, with the relaxation time $\tau_r \simeq N_\text{col}/\Gamma_\text{el}$, which is the time needed to restore \textit{kinetic} equilibrium. Here, $\Gamma_\text{el}$ is the elastic scattering rate and $N_\text{col} \simeq \max\{1, m_\chi/T\}$ is the number of collisions needed to redistribute any temperature differences between the two sectors.
For a given elastic scattering cross-section $\sigma_{\chi e^\pm \leftrightarrow \chi e^\pm}$, we thus estimate the kinetic decoupling temperature via the relation, $1/\tau_\text{el}(\Tkd) \sim H(\Tkd)$, or more explicitly ($\Gamma_\text{el} = n_e \langle \sigma_{\chi e^\pm \leftrightarrow \chi e^\pm} v \rangle$)
\begin{align}
    \frac{n_e (\Tkd) \langle \sigma_{\chi e^\pm \leftrightarrow \chi e^\pm} v \rangle(\Tkd)}{N_\text{col} (\Tkd)} \sim H(\Tkd)\eqsp.
    \label{eq:def_Tkin}
\end{align}
This expression involves the number density $n_e$ of electrons and positrons, as well as the thermally averaged cross-section (assuming Maxwell-Boltzmann statistics)
\begin{align}
\langle \sigma_{\chi e^\pm \leftrightarrow \chi e^\pm} v \rangle \simeq \frac{1}{2T m_\chi^2 m_e^2 K_2(m_e/T) K_2(m_\chi/T)} \int_{(m_\chi + m_e)^2}^\infty \text{d} s\; \sigma_{\chi e^\pm \leftrightarrow \chi e^\pm}(s) p_{\chi e}(s)^2\sqrt{s}K_1(\sqrt{s}/T)
\label{eq:sig_v_kd}
\end{align}
with
\begin{align}
    p_{\chi e}(s) = \frac{[s - (m_\chi + m_e)^2]^{1/2} [s - (m_\chi - m_e)^2]^{1/2}}{2\sqrt{s}}\eqsp.
\end{align}
At $T=\Tkd$, the electrons/positrons may already be non-relativistic, in which case the baryon-asymmetry of the Universe becomes important for evaluating $n_e$. For this reason, we parameterize the corresponding number density as
\begin{align}
n_e(T) \simeq \max \big\{ n_e^\text{eq}(T), n_e^\text{asym}(T)\big\}
\end{align}
with
\begin{align}
n_e^\text{eq}(T) = g_e \int \frac{\text{d}^3 p}{(2\pi)^3} \frac{1}{\exp(E/T) + 1} \qquad \text{and}\qquad n_e^\text{asym}(T) \simeq \Big(Y_p +2Y_ {{}^4\text{He}}\Big) n_b(T)\eqsp.
\end{align}
Here, $g_e = 4$, $n_b = 2\zeta(3) \eta T^3/\pi^2$ is the baryon number density with the baryon-to-photon ratio $\eta$, and $Y_p = n_p/n_b$ ($Y_{{}^4\text{He}} = n_ {{}^4\text{He}}/n_b$) is the abundance of protons (helium-4) with the corresponding number density $n_p$ ($n_{{}^4\text{He}}$). For large temperatures, we therefore determine the density of electrons and positrons via their thermal distributions, while for low temperatures only electrons remain ($n_{e^+} \simeq 0$) with a density determined by the baryon asymmetry of the Universe. Using this expression, it is thus possible to evaluate $\Tkd$ and consequently $T_\chi$ and $\langle \sigma_\text{ann} v \rangle$, which ultimately enables the evaluation of the source terms in eq.~\eqref{eq:source_terms}.

At this point, let us note that the general form of $\sigma_{\chi e^\pm \leftrightarrow \chi e^\pm}$ and thus the corresponding value of $\Tkd$ is naturally model dependent. In order to account for this fact, while also being able to present more general (model-independent) results, in the following (cf.~sec.~\ref{sec:results}) we discuss both fixed values of $\Tkd$ (analogues to \cite{Depta:2019lbe}) as well as dynamic values of $\Tkd$ based on eq.~\eqref{eq:def_Tkin} and the three benchmark models shown in tab.~\ref{tab:models}. The corresponding expressions for $\sigma_{\chi e^\pm \leftrightarrow \chi e^\pm}$ are summarized in tab.~\ref{tab:sig_xexe}.

\begin{table}[t]
    \centering
    \begin{tabular}{c|c}
        ID & $\sigma_{\chi e^\pm \leftrightarrow \chi e^\pm}(s) \times s^2 m_R^4/(4 \pi \gamma_d \gamma_v)$ \\ \hline \\
        (1) & $4m_\chi^2\left[s^2-2s(m_\chi^2-3m_e^2) + (m_e^2 -m_\chi^2)^2 \right]$ \\ \\
        (2) & $4s^3 -10 s^2 (m_\chi^2+m_e^2) + s(9m_\chi^4 +22 m_\chi^2 m_e^2 + 9m_e^4)  -4(m_\chi^4 -m_e^4)(m_\chi^2-m_e^2) +(m_\chi^2-m_e^2)^4/s $\\ \\
        (3) & $18 \big[m_\chi^4 (m_e^2+s) -2 m_\chi^2(m_e^2-s)^2 + (m_e^4-s^2)(m_e^2-s) \big]$
    \end{tabular}
    \caption{The cross-section $\sigma_{\chi e^\pm \leftrightarrow \chi e^\pm}$ for the three benchmark models shown in tab.~\ref{tab:models} in terms of the parameters $\gamma_v$ and $\gamma_d$ as well as the masses $m_\chi \in \{m_\varphi, m_\psi \}$ and $m_R \in \{ m_\Phi, m_{A'}\}$.}
    \label{tab:sig_xexe}
\end{table}

\begin{figure}
\centering
\includegraphics[width=0.49\linewidth]{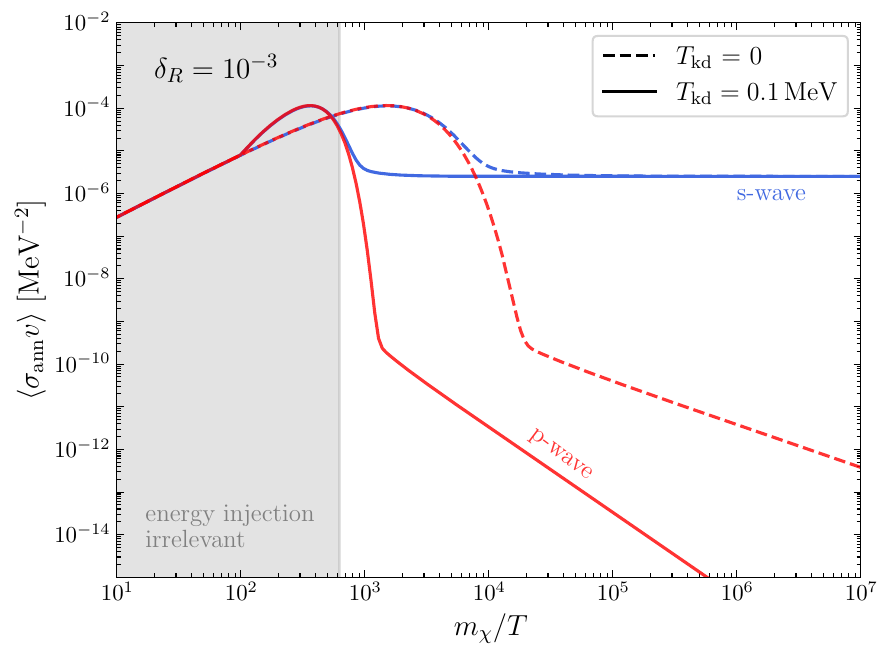}
\includegraphics[width=0.49\linewidth]{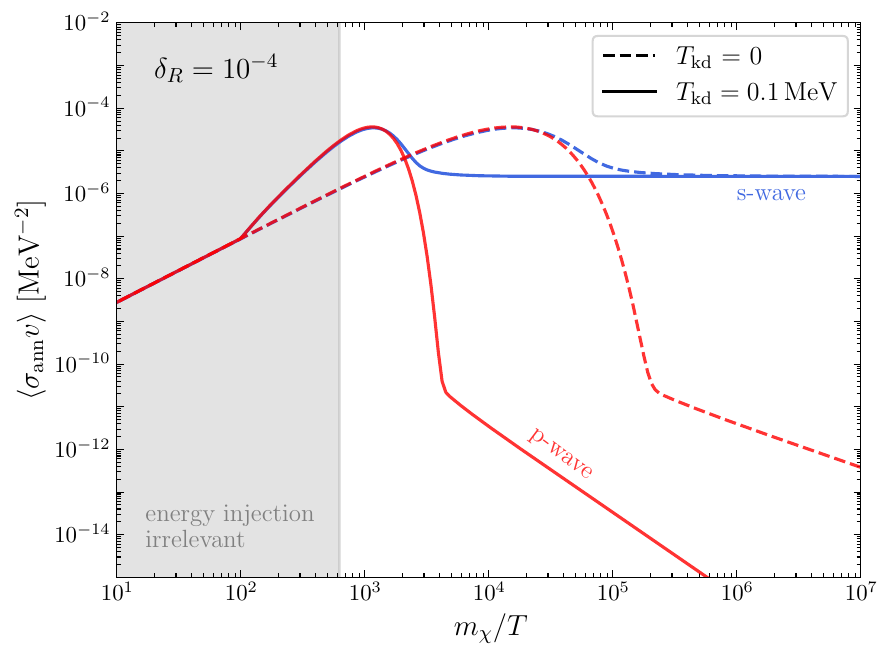}
\caption{The thermally averaged cross-section as a function of temperature for $s$-wave (blue) and $p$-wave (red) annihilations for $\dm = 10^{-3}$ (left) as well as $\dm = 10^{-4}$ (right). Here, we choose $m_\chi = 10\,\mathrm{MeV}$, $\gamma_d = 10^{-3}$ and $\gamma_v= 10 \times \dm^2$, with the latter choice ensuring that $a$ ($b$) is constant for $s$-wave ($p$-wave) annihilations. We compare the evolution of the cross-section in case of kinetic decoupling at $\Tkd = 0.1\,\mathrm{MeV}$ (solid), as well as without kinetic decoupling, i.e.~$\Tkd = 0$ (dashed).}
    \label{fig:sigv}
\end{figure}

Finally, to demonstrate the impact of $\Tkd$ on the overall results, in fig.~\ref{fig:sigv} we show the thermally averaged cross-section as a function of $m_\chi/T$ for $s$-wave (blue) and $p$-wave annihilations (red), $\dm = 10^{-3}$ (left) and $\dm = 10^{-4}$ (right), as well as $\Tkd = 0.1\,\mathrm{MeV}$ (solid) and $\Tkd = 0$ (dashed). For this plot, we fix $m_\chi = 10\,\mathrm{MeV}$, $\gamma_d = 10^{-3}$ and $\gamma_v = 10\times \dm^2$, with the latter choice ensuring that $a$ ($b$) is constant for $s$-wave ($p$-wave) annihilations. Moreover, the white region indicates the range of temperatures for which photodisintegration is relevant, i.e. the region for which $T < T_\text{max} \sim \mathrm{O}(1)\,\mathrm{keV}$~\cite{Hufnagel:2018bjp}. Here, we choose $T_\text{max} \approx 3\,\mathrm{keV}$ in accordance with the implementation in \acropolis.\footnote{For higher temperatures, the photons resulting from the electromagnetic cascade have too little energy to dissociate any light elements (for more information, see~\cite{Hufnagel:2018bjp}).} In general, we find that larger values of $\Tkd$ shift the resonance peak to higher temperatures -- outside the region relevant for photodisintegration --, while simultaneously decreasing the overall width of the peak. Lower values of $\Tkd$ therefore enhance the effect of the resonance and we can already anticipate that the bounds from resonantly-enhanced annihilations will differ more strongly from the vanilla scenario with $\langle \sigma_\text{ann} v \rangle = a + 6b/x$ for smaller values of $\Tkd$. We will come back to this discussion once we discuss the actual limits in sec.~\ref{sec:results}.


\section{Implementation in \acropolis}
\label{sec:implementation}

\subsection{The class \texttt{acropolis.ext.models.ResonanceModel}}

\noindent Up until v1.2.2, \acropolis included two different types of models: \texttt{acropolis.models.DecayModel}, which can be used to calculate photodisintegration constraints for unstable DS particles decaying into electromagnetic final states, as well as \texttt{acropolis.models.AnnihilationModel}, which allows for calculating bounds on vanilla $s$- and/or $p$-wave annihilations. However, none of these two models can handle the resonantly-enhanced annihilations described in this work. Therefore, based on the results of the previous section, in v1.3.0, we implement and make available a new model \texttt{acropolis.ext.models.ResonanceModel}, which inherits directly from \texttt{AnnihilationModel},\footnote{\texttt{AnnihilationModel} already implements source terms of the form~\eqref{eq:source_terms} albeit for the special case $\langle \sigma v \rangle = a + b\langle v^2 \rangle$ with some constants $a$ and $b$. Thus, by using this model as a base class, we only have to modify the calculation of the thermally averaged annihilation cross-section, while all the other logic can remain unchanged.} yet features a different set of input parameters and overrides the function \texttt{sigma\_v} in order to incorporate the modified calculation of the annihilation cross-section based on eq.~\eqref{eq:sigv_general}. Specifically, the class constructor of this new model takes the following arguments
\begin{itemize}
    \item \texttt{mchi}: The mass $m_\chi$ of the DM particle in MeV
    \item \texttt{delta}: The parameter $\dm$ describing the mass splitting between the DM particle and the mediator as defined in eq.~\eqref{eq:def_eps}
    \item \texttt{gammad}: The coupling constant $\gamma_d$ encoding the interaction between the mediator and the DM particle as defined in eq.~\eqref{eq:gammas}
    \item \texttt{gammav}: The coupling constant $\gamma_v$ encoding the interaction between the mediator and the SM particle as defined in eq.~\eqref{eq:gammas}
    \item \texttt{nd}: The parameter $n_d$ discriminating between $s$-wave ($n_d = 0$) and $p$-wave annihilations ($n_d = 1$) as defined in eq.~\eqref{eq:gammas}
    \item \texttt{tempkd}: The SM temperature $\Tkd$ in MeV at which the DM particles decouple kinetically from the SM heat bath as defined in eq.~\eqref{eq:def_Tkin}
    \item \texttt{S}: The symmetry factor $S$ entering the cross-section in eq.~\eqref{eq:sig_ann} \texttt{[default = 1]}
    \item \texttt{omegah2}: The relic density of the DM particles \texttt{[default = 0.12]}
\end{itemize}
Given these parameters, the class \texttt{ResonanceModel} can be used like any other model that is already part of \acropolis (for more details on how to use \acropolis, see \cite{Depta:2020mhj}), e.g.~via the code
\enlargethispage*{\baselineskip}
\begin{lstlisting}
# ext.models
from acropolis.ext.models import ResonanceModel

# Initialize the model
model = ResonanceModel(
    mchi   = 10. ,
    delta  = 1e-2,
    gammad = 1e-5,
    gammav = 1e-3,
    nd     = 0   ,
    tempkd = 1e-2
)

# Run photodisintegration
Yf = model.run_disintegration()
\end{lstlisting}
for a model with $m_\chi = 10\,\mathrm{MeV}$, $\dm = 10^{-2}$, $\gamma_d = 10^{-5}$, $\gamma_v = 10^{-3}$, $n_d = 0$, and $\Tkd = 10\,\mathrm{keV}$.

However, note that according to eq.~\eqref{eq:def_Tkin}, $\Tkd$ is strictly speaking not an independent quantity, but rather a function of other model parameters. To account for this fact, while also allowing for a more model-independent analysis, we implemented the model in such a way that \texttt{tempkd} can either be fixed, i.e.~constant (as above), or \textit{any} function with signature\footnote{More precisely, we allow any object with the given signature and \texttt{callable(tempkd) == True}.}
\begin{lstlisting}
def my_tempkd_func(mchi, delta, gammad, gammav, nd, S, ii):
    [...]
\end{lstlisting}
by setting \texttt{tempkd = my\_tempkd\_func}.
Here, the first six parameters are identical to the first six ones discussed above, while the parameter \texttt{ii} is an instance of \texttt{acropolis.input.InputInterface}, which allows access to parameters like e.g.~the baryon-to-photon ratio $\eta$ or the Hubble rate $H$ from within the function (as required e.g.~to implement eq.~\eqref{eq:def_Tkin}). While it is possible to implement any such function from scratch, we also provide a reference function
\begin{lstlisting}
estimate_tempkd_ee(mchi, delta, gammad, gammav, nd, S, ii, sigma_ee)
\end{lstlisting}
in \texttt{acropolis.ext.models}, which implements the calculation of $\Tkd$ according to eqs.~\eqref{eq:def_Tkin} and \eqref{eq:sig_v_kd} for a given (model-dependent) cross-section $\sigma_{\chi e^\pm \leftrightarrow \chi e^\pm}(s)$. The latter is provided via the additional parameter \texttt{sigma\_ee}, which can be any function with signature
\begin{lstlisting}
def my_sigma_ee_func(s, mchi, delta, gammad, gammav):
    [...]
\end{lstlisting}
Here, the parameter \texttt{s} encodes the center-of-mass energy $s$. Consequently, for a given cross-section $\sigma_{\chi e^\pm \leftrightarrow \chi e^\pm}(s)$, one way of creating a function \texttt{my\_tempkd\_func} that implements eqs.~\eqref{eq:def_Tkin} and \eqref{eq:sig_v_kd}, while also being compatible with the parameter \texttt{tempkd} in the constructor of \texttt{ResonanceModel} is via
\begin{lstlisting}
# functools
from functools import partial

# ext.models
from acropolis.ext.models import estimate_tempkd_ee

my_tempkd_func = partial(estimate_tempkd_ee, sigma_ee=my_sigma_ee_func)
\end{lstlisting}

Finally, let us note that for the calculation of $\langle \sigma_\text{ann} v\rangle$ entering the source terms in eq.~\eqref{eq:source_terms},\footnote{See \texttt{acropolis.ext.models.ResonanceModel.sigma\_v}, which is inherited from \texttt{AnnihilationModel}.} by default, we perform the full integral from eq.~\eqref{eq:sigv_general} numerically. The corresponding implementation can be found in the function \texttt{ResonanceModel.\_sigma\_v\_full}. However, for convenience, we have also implemented the two approximate expressions for $\langle \sigma_\text{ann} v \rangle^\text{res}$ from eq.~\eqref{eq:sigma_res} and $\langle \sigma_\text{ann} v \rangle^\text{non-res}$ from eq.~\eqref{eq:sigma_non_res} in the functions \texttt{ResonanceModel.\_sigma\_v\_res} and \texttt{ResonanceModel.\_sigma\_v\_non\_res}, respectively.

\subsection{Implementing benchmark scenarios}
Any benchmark model, including the ones from tab.~\ref{tab:models}, can thus be studied via \texttt{ResonanceModel} by fixing \texttt{nd}, \texttt{S}, and setting \texttt{tempkd} to an appropriate function, which can either be implemented from scratch or derived from \texttt{estimate\_tempkd\_ee} by providing the corresponding scattering cross-section. For the benchmark models in tab~\ref{tab:models}, we have already implemented the relevant cross-sections from tab.~\ref{tab:sig_xexe} in \texttt{acropolis.ext.benchmarks}; specifically in the functions \texttt{sigma\_ee\_bx} with \texttt{x = 1, 2, 3}. In this module, we have -- for convenience -- further implemented 'subclasses' of \texttt{ResonanceModel} for the three benchmark models defined in tab.~\ref{tab:models}, namely,
\begin{lstlisting}[backgroundcolor=\color{white}]
acropolis.ext.benchmarks.BenchmarkModel1
acropolis.ext.benchmarks.BenchmarkModel2
acropolis.ext.benchmarks.BenchmarkModel3
\end{lstlisting}
corresponding to the models $(1)$, $(2)$, and $(3)$, respectively.
These models can be initiated like \texttt{ResonanceModel}, but without the need to specify \texttt{nd}, \texttt{S}, and \texttt{tempkd}, as these parameters are already set accordingly. 

Moreover, by using the provided tools, it is further possible to create additional benchmark models. This can be done by again utilizing \texttt{functools.partial}, i.e.~via
\begin{lstlisting}
# functools
from functools import partial

# ext.models
from acropolis.ext.models import ResonanceModel

MyBenchmarkModel = partial(
    ResonanceModel, nd = [...], S = [...], tempkd = [...]
)
\end{lstlisting}

\subsection{Running parameter scans}
Bounds in any two-parameter plane can be derived using \acropolis's build-in scanning framework, i.e. the class \texttt{acropolis.scans.BufferedScanner}, with either the general \texttt{ResonanceModel}, or any of the specialized benchmark models (cf.~\cite{Depta:2020mhj} for more information). For example, the following code can be used to run a \texttt{N}x\texttt{N} scan for benchmark model $(3)$ with $\dm = 10^{-2}$, $\gamma_d = 10^{-3}$, as well as $m_\chi \in \{1, 10^{3}\}\,\mathrm{MeV}$ and $\gamma_v \in \{ 10^{-14}, 10^{-2} \}$
\begin{lstlisting}
# scans
from acropolis.scans import BufferedScanner, ScanParameter
# ext.benchmarks
from acropolis.ext.benchmarks import BenchmarkModel3

scan_result = BufferedScanner( BenchmarkModel3,
                               mchi   = ScanParameter(  0,  3, N),
                               delta  = 1e-2,
                               gammad = 1e-3,
                               gammav = ScanParameter(-14, -2, N),
                             ).perform_scan(cores=-1)
\end{lstlisting}
Scans of this type are computationally expensive and should be run on a machine with many CPU cores. In principle, \acropolis allows speeding up the calculation under certain conditions by setting \texttt{fast=True} for one of the two \texttt{ScanParameter} objects, which can reduce computation time by several orders of magnitude. For \texttt{AnnihilationModel} and thus \texttt{ResonanceModel} this is possible for any parameter (usually some coupling) $g$ that \textit{(i)} only enters the annihilation cross-section, and \textit{(ii)} fulfills $\langle \sigma_\text{ann} v \rangle \propto g$. However, while $\Gamma_v(p) \Gamma_d(p_f) \propto \gamma_v \gamma_d$, we still have $\langle \sigma_\text{ann} v \rangle \,\cancel\propto\, \gamma_v \gamma_d$, since $\gamma_v$ and $\gamma_d$ also enter the total width $\Gamma(p)$ in the denominator (cf.~eq.~\eqref{eq:sigma_fp}). Moreover, for more model-dependent scenarios, $\gamma_v$ and $\gamma_d$ also enter the calculation of $\Tkd$. Consequently, it is -- in general -- not possible to speed up scans involving \texttt{ResonanceModel} by setting \texttt{fast=True}. For this reason, we recommend to run the calculation on a cluster or a machine with many cores.

However, for fixed $\Tkd$, in the limit where one coupling is much larger than the other, the total decay width is roughly proportional to the larger coupling. As a result, $\langle \sigma_\text{ann} v \rangle$ is proportional to the smaller coupling ($\gamma_v$ in our case, see below). Scanning over this coupling thus warrants the use of \texttt{fast=True}, speeding up the calculation significantly.
The bounds for fixed $\Tkd$ and $\gamma_d$ presented below were therefore derived using \texttt{fast=True}. However, we additionally verify this choice by checking that $\gamma_d \gg \gamma_v$ holds true all along the exclusion lines.

\section{Results and Discussion}
\label{sec:results}
\subsection{Constraints on $s$-wave annihilations}

\begin{figure}[t]
    \centering
    \includegraphics[width=0.495\linewidth]{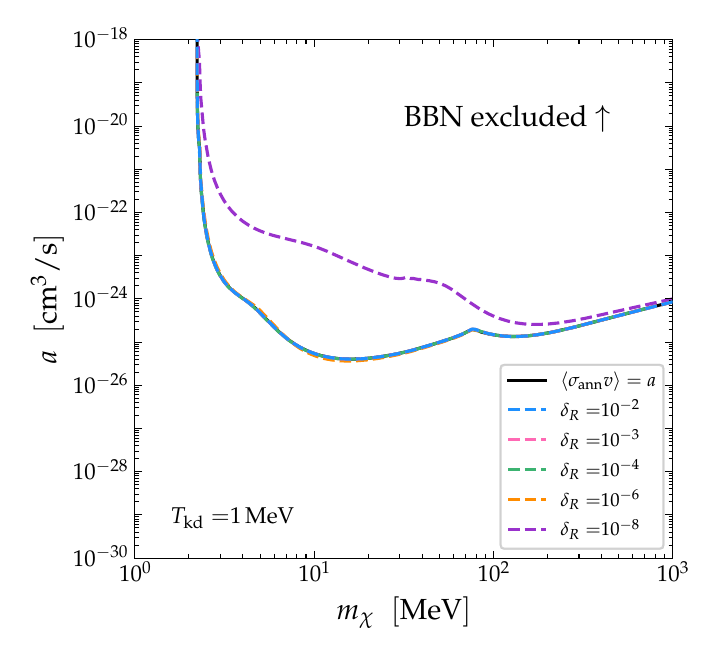}
    \includegraphics[width=0.495\linewidth]{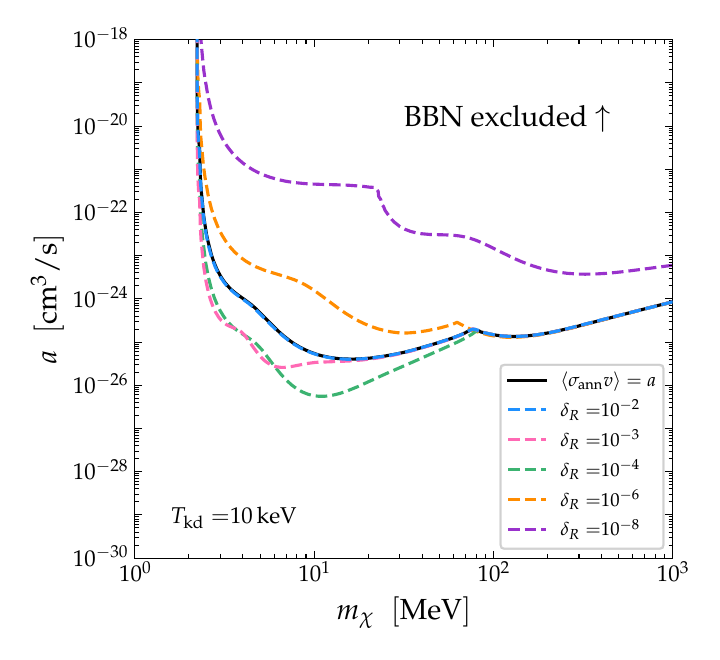}
    \includegraphics[width=0.495\linewidth]{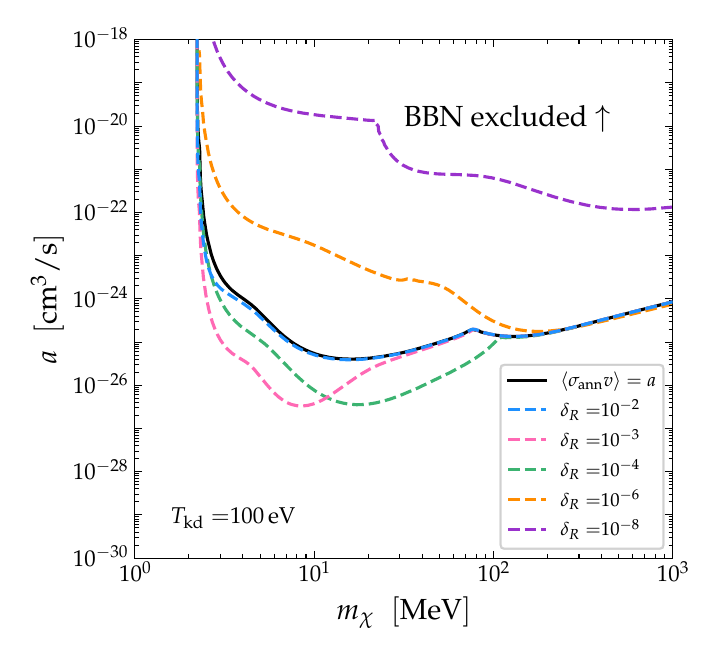}
    \caption{BBN constraints from photodisintegration at 95\% C.L. on resonant DM annihilations for $\gamma_d = 10^{-3}$, $n_d = 0$, different values of $\dm \in \{10^{-2}, 10^{-3}, 10^{-4}, 10^{-6}, 10^{-8}\}$ (dashed, different colors), as well as different values of $\Tkd \in \{1, 10^{-2}, 10^{-4}\}\,\mathrm{MeV}$ (different panels) in the $a-m_\chi$ parameter plane. For comparison, we also show the constraints that are obtained for non-resonant annihilations of NR DM particles (solid, black), i.e. for $\langle \sigma_\text{ann} v \rangle = a$.}
    \label{fig:L=0results}
\end{figure}

In this section, we first present the constraints from photodisintegration for $s$-wave annihilations, i.e. for $n_d = 0$.\footnote{For the complementary bound on the mass of thermal relics, see e.g.~\cite{Depta:2019lbe}.} In the absence of any resonance effects, limits on $s$-wave annihilations of NR DM particles are usually given in terms of the parameter $a$ with $\langle \sigma_\text{ann} v \rangle \simeq a$. While such a parametrization is not applicable in our scenario, in the non-resonant regime we can utilize eq.~\eqref{eq:defa} to identify
\begin{align}
    a = \frac{2S\pi\gamma_v \gamma_d}{m_\chi^2} \frac{(2+\dm)^2}{\dm^2} \overset{\dm \ll 1}{\simeq} \frac{8S\pi\gamma_v \gamma_d}{m_\chi^2 \dm^2}\eqsp.
\end{align}
Consequently, by presenting our results in terms of this parameter $a$ -- which can be exchanged for one of the couplings, i.e.~$\gamma_v$ (see below) --, we enable a meaningful comparison with previously published, non-resonant constraints \cite{Depta:2019lbe}. By fixing $\gamma_d = 10^{-3}$, $n_d = 0$,\footnote{The parameter $S$ is irrelevant, as it can be absorbed into $a$.} and scanning over $\gamma_v$ and $m_\chi$, in fig.~\ref{fig:L=0results}, we show the resulting constraints in the $a-m_\chi$ parameter plane for different values of the mass splitting $\dm \in \{10^{-2}, 10^{-3}, 10^{-4}, 10^{-6}, 10^{-8}\}$ (dashed, different colors), as well as for fixed values of $\Tkd \in \{1, 10^{-2}, 10^{-4} \}\,\mathrm{MeV}$ (different panels), which have been obtained by running \acropolis with \texttt{ResonanceModel}. For comparison, we also indicate the constraints that are obtained in the absence of any resonance (solid, black), i.e.~for $\langle \sigma_\text{ann} v \rangle = a$ everywhere, which we obtain by using \texttt{AnnihilationModel} instead.\footnote{These results correspond to the ones obtained in \cite{Depta:2019lbe}, albeit with updated values for the observed nuclear abundances.} For different choices of $\gamma_d$ we refer the reader to appendix~\ref{app:gammad}.

Just like in the case of vanilla $s$-wave annihilations, we find that it is not possible for photodisintegration to constrain DM particles with masses below $m_\chi \sim 2.22\,\mathrm{MeV}$, corresponding to the binding energy of deuterium. This is because, for smaller masses, the annihilation products with energy $E = m_\chi$ are not energetic enough to dissociate deuterium or any other relevant nuclei, meaning that the light-element abundances remain unaltered.

For larger masses, however, i.e.~once photodisintegration reactions become relevant, we find that the constraints from resonantly-enhanced annihilations become relevant and also potentially different from the ones obtained with $\langle \sigma_\text{ann} v \rangle = a$ (the ``vanilla" scenario). To quantify this difference, let us note that we generally expect the constraints to differ from the vanilla scenario if the resonance peak is pushed below $T_\text{max} = \mathcal{O}(1)\,\mathrm{keV}$, i.e.~into the temperature range relevant for photodisintegration (cf.~fig.~\ref{fig:sigv}). Assuming $\gamma_d \gg \gamma_v$, which is true for the parameter combinations presented in fig.~\ref{fig:L=0results}, we find the position of the resonance peak to be at $m_\chi/T_\chi \sim 3/(2\dm)$ (cf.~eq.~\eqref{eq:sigma_res}) with the corresponding annihilation cross-section $\langle \sigma_\text{ann} v \rangle = \mathcal{O}(1) a \sqrt{\dm}/\gamma_d$. 
Hence, $\langle \sigma_\text{ann} v \rangle \gtrsim a$ ($\langle \sigma_\text{ann} v \rangle \lesssim a$) for $\dm \gtrsim \mathcal{O}(1)\gamma_d^2$ ($\dm \lesssim \mathcal{O}(1)\gamma_d^2$). This directly translates to an expected strengthening (weakening) of the constraints compared to the ones with $\langle \sigma_\text{ann} v \rangle = a $ everywhere. However, as mentioned above, this argument only holds true if the resonance contribution peaks within the relevant temperature range, i.e.~for $T_\text{peak} \lesssim T_\text{max}$ with $m_\chi/T_\chi(T_\text{peak}) = 3/(2\dm)$. Enforcing this condition, we \textit{quantitatively} find $m_\chi \lesssim \mathcal{O}(1)T_\text{max}/\dm$ for $T_\text{peak} > \Tkd$ as well as $m_\chi \lesssim \mathcal{O}(1) T_\text{max}^2/(\Tkd \dm)$ for $T_\text{peak} \geq \Tkd$. Consequently, we expect the bounds to differ only over a certain range of masses, with this range becoming larger for smaller values of $\dm$. The same applies for smaller values of $\Tkd$ when $T_\text{peak} \geq \Tkd$.

For given values of $\gamma_d$ and $\Tkd$, when reducing $\dm$, the bounds therefore differ from the vanilla ones for a larger range of masses, while being stronger for values of $\dm$ above some critical value $\dm^\text{crit}$. However, the value of $\dm^\text{crit}$ is usually even larger than $\mathcal{O}(1)\gamma_d^2$ (as expected from the argument above). To understand this, let us note that photodisintegration is most sensitive to temperatures slightly below $T_\text{max}$, since $S^{(0)}(T) \propto n_\chi(T)^2 \propto T^6$, meaning that the amount of injected energy drops sharply with temperature. However, for $T \gg T_\text{peak}$, $\langle \sigma_\text{ann} v \rangle \propto 1/T^{3/2}$, meaning that if the peak is pushed to small enough temperatures, we might have $\langle \sigma_\text{ann} v \rangle < a$ at $T = T_\text{max}$, while still $\langle \sigma_\text{ann} v \rangle > a$ at $T = T_\text{peak}$ (cf.~the dashed blue line in the right panel of fig.~\ref{fig:sigv}). Therefore, in general $\dm^\text{crit} > \mathcal{O}(1)\gamma_d^2$ with $\gamma_d^2 = 10^{-6}$ for the given choice of parameters.



\begin{figure}
\centering
\includegraphics[width=0.495\linewidth]{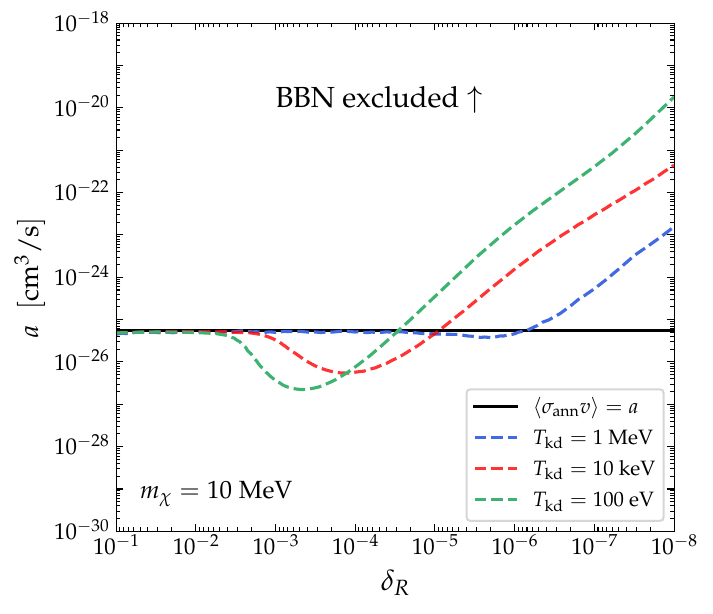}
\includegraphics[width=0.495\linewidth]{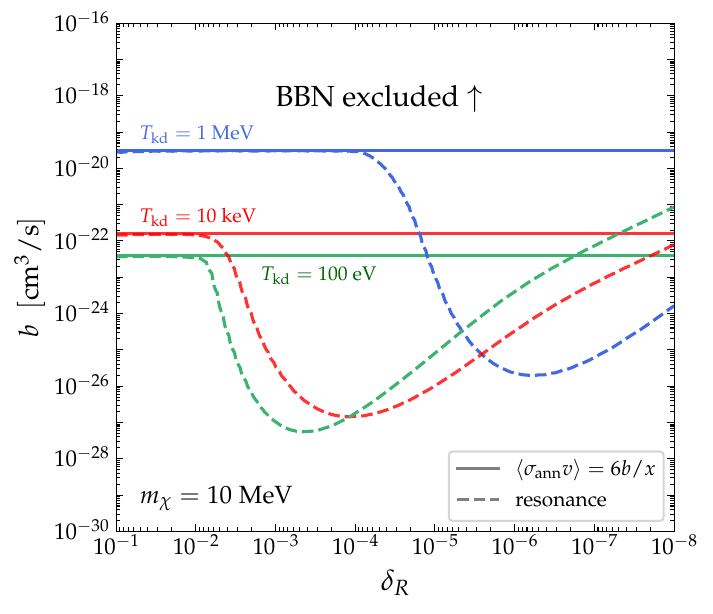}
\caption{BBN constraints from photodisintegration at 95\% C.L. for resonantly enhanced $s$-wave (left) or $p$-wave (right) annihilations as a function of $\dm$ for $m_\chi = 10\,\mathrm{MeV}$ (dashed), and different decoupling temperatures (different colors). Also shown are the bounds on vanilla $s$- and $p$-wave annihilations (solid). Note the inverted $x$-axis.}
\label{fig:delta}
\end{figure}

Overall, the behaviour described above is reflected in fig.~\ref{fig:L=0results}: Taking $\Tkd = 100\,\mathrm{eV}$ as an example, we find that the point at which a given colored line merges with the black line gets pushed to larger values of $m_\chi$ for smaller values of $\dm$. Also, while we see an initial improvement of the bounds for $\dm \ll 10^{-6} = \gamma_d^2$, the bounds do indeed become weaker than the ones with $\langle \sigma_\text{ann} v \rangle = a$ for smaller values of $\dm$. The actual value of $\dm^\text{crit}$ thus lies somewhere between $\dm = 10^{-4}$ and $\dm = 10^{-6}$. Similar results are also obtained for different values of $\Tkd$. To further illustrate this point, in fig.~\ref{fig:delta} we additionally show the resulting bounds in the $a-\dm$ (left) and $b-\dm$ (right, cf.~sec.~\ref{sec:pwave_results}) parameter space for fixed DM mass $m_\chi = 10$ MeV, different values of $\Tkd$ (different colors), as well as with (dashed) and without (solid) resonance effects. Based on this figure, we can identify approximate values for $\dm^\text{crit}$, which turn out to be $\dm^\text{crit} \sim 10^{-6}, 10^{-5}, 7\times10^{-4} $ for $\Tkd = 1\,\mathrm{MeV}, 10\,\mathrm{keV}, 100\,\mathrm{eV}$, respectively.

In order to strengthen the constraints compared to the scenario with $\langle \sigma_\text{ann} v \rangle = a$, it is therefore necessary to tune $\Tkd$ and $\dm$, accordingly. However, such a tuning might not always be possible for concrete models with dynamically calculated values of $\Tkd$, as we will see below.

\begin{figure}[t]
\centering
\includegraphics[width=0.495\linewidth]{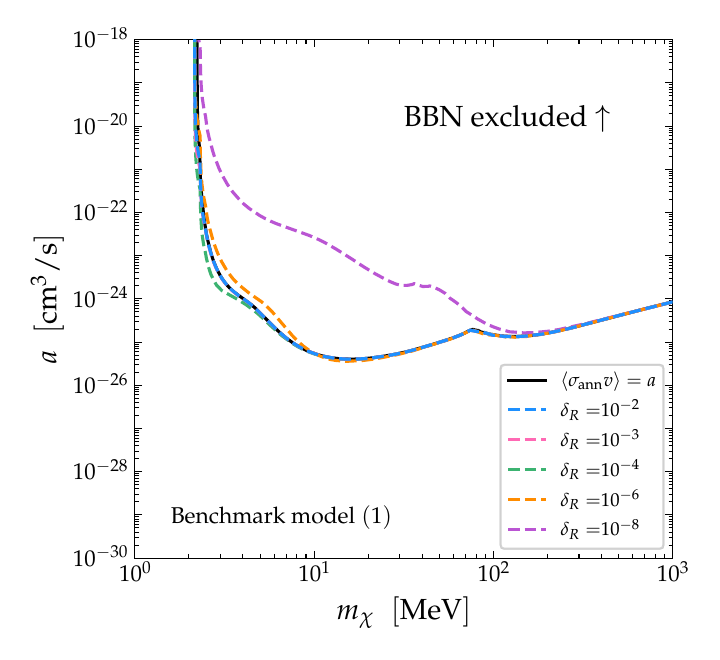}
\includegraphics[width=0.495\linewidth]{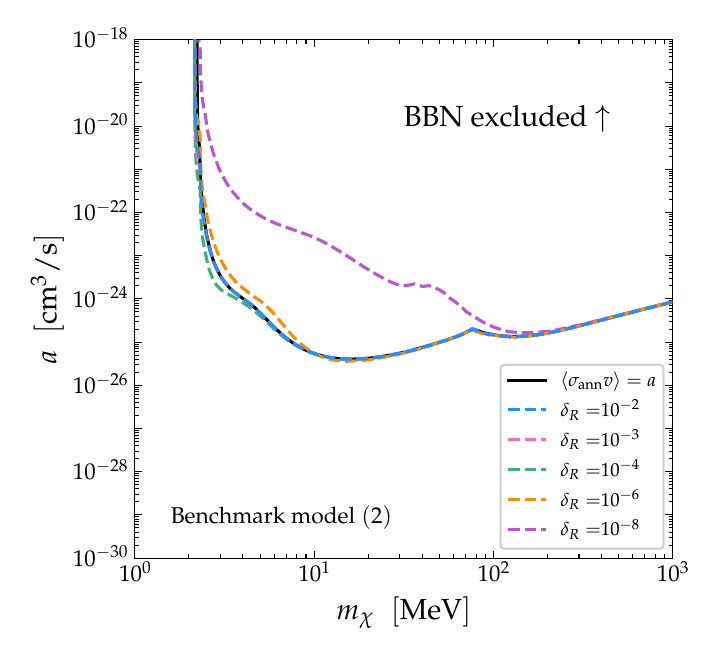}
\caption{BBN constraints from photodisintegration at 95\% C.L. on resonant DM annihilations for $\gamma_d = 10^{-3}$, $n_d = 0$, different values of $\dm \in \{10^{-2}, 10^{-3}, 10^{-4}, 10^{-6}, 10^{-8}\}$ (dashed, different colors), and dynamically calculated values of $\Tkd$ according to eq.~\eqref{eq:def_Tkin} for the benchmark models $(1)$ (left) and $(2)$ (right) of tab.~\ref{tab:models}.}
\label{fig:L=0results2}
\end{figure}

In addition to fixed values of $\Tkd$, in fig.~\ref{fig:L=0results2}, we further show the resulting constraints for dynamically calculated values of $\Tkd$ according to eq.~\eqref{eq:def_Tkin} corresponding to the two benchmark models $(1)$ (left) and $(2)$ (right) from tab.~\ref{tab:models}, which have been obtained by running \acropolis with \texttt{BenchmarkModel1} and \texttt{BenchmarkModel2}, respectively. Most notably, in this case, the resulting constraints are much more similar to the ones obtained with $\langle \sigma_\text{ann} v \rangle = a$, at least for large values of $\dm$. In fact, when calculating the kinetic decoupling temperature in the given benchmark models, we consistently find comparatively large kinetic decoupling temperatures, $\Tkd \gtrsim 1\,\mathrm{MeV}$, for all parts of parameter space. Due to this, the resulting constraints strongly resemble those in the top left panel of fig.~\ref{fig:L=0results}. As it turns out, pushing $\Tkd$ to values larger than $1\,\mathrm{MeV}$ does not lead to an appreciable change in the limits, since $\Tkd > 1\,\mathrm{MeV}$ is already much larger than $T_\text{max}$, meaning that kinetic decoupling happens anyhow outside the photodisintegration window.
Significant differences between the two scenarios are only obtained for small values of $\dm \sim 10^{-8}$, in which case the constraints weaken. Overall, we therefore conclude that -- at least for the benchmark models discussed in this work -- it is difficult to strengthen the photodisintegration constraints within a minimal realistic scenario by resonantly-enhancing the annihilation cross-section. However, if additional interactions are present between the DS and SM states, it would in principle be possible to lower $\Tkd$, which would lead to stronger constraints.

\subsection{Constraints on $p$-wave annihilations}
\label{sec:pwave_results}
\begin{figure}[t]
    \centering
    \includegraphics[width=0.495\linewidth]{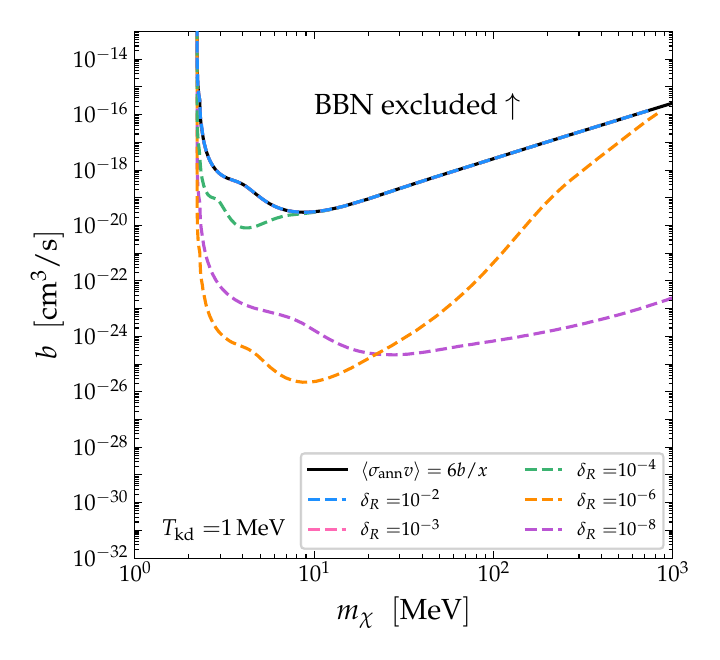}
    \includegraphics[width=0.495\linewidth]{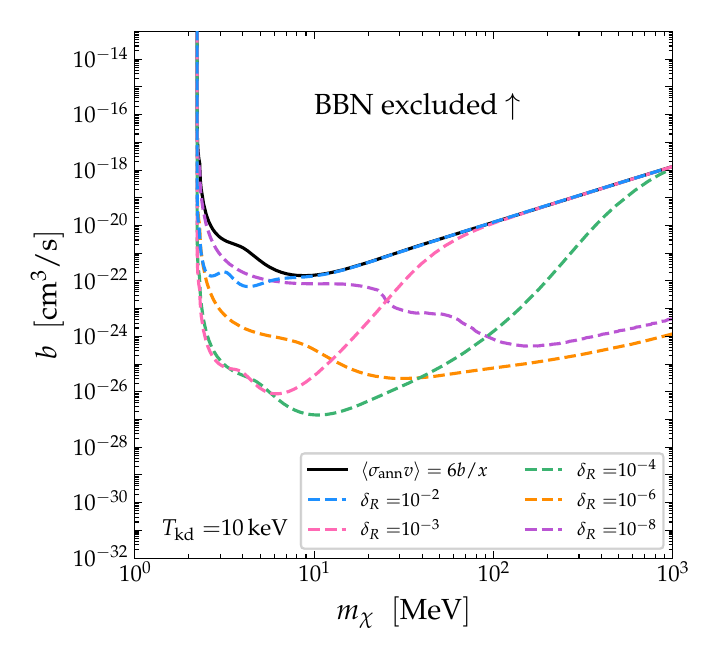}
    \includegraphics[width=0.495\linewidth]{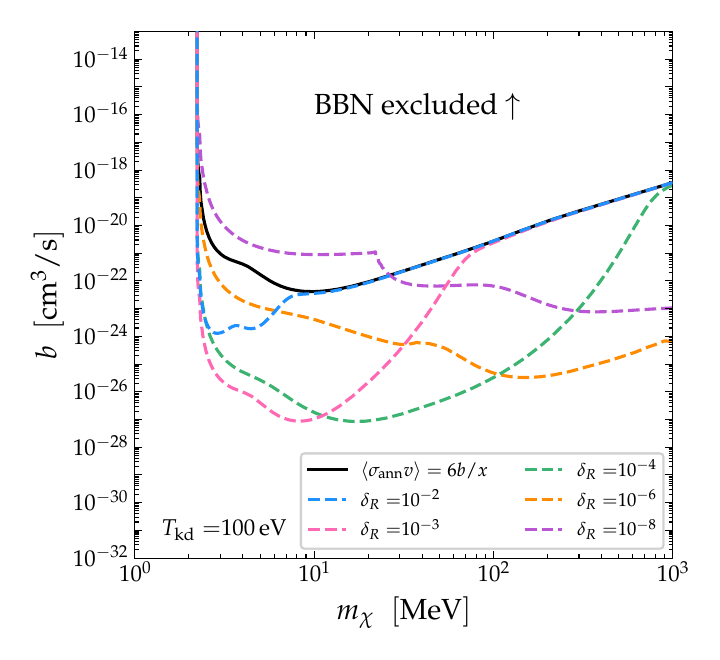}
    \caption{BBN constraints from photodisintegration at 95\% C.L. on resonant DM annihilations for $\gamma_d = 10^{-3}$, $n_d = 1$, different values of $\dm \in \{10^{-2}, 10^{-3}, 10^{-4}, 10^{-6}, 10^{-8}\}$ (dashed, different colors), as well as different values of $\Tkd \in \{1, 10^{-2}, 10^{-4}\}\,\mathrm{MeV}$ (different panels) in the $b-m_\chi$ parameter plane. For comparison, we also show the constraints that are obtained for non-resonant $p$-wave annihilations of NR DM particles (solid, black), i.e.~for $\langle \sigma_\text{ann} v \rangle = 6b/x$.}
    \label{fig:L=1results}
\end{figure}

In this section, we present our results for $p$-wave annihilations, i.e.~for $n_d = 1$. Following the strategy from the previous section, we present our results in terms of the parameter
\begin{align}
b = \frac{S\pi\gamma_v \gamma_d}{m_\chi^2} \frac{(2+\dm)^2}{2\dm^2} \overset{\dm \ll 1}{\simeq} \frac{2S\pi\gamma_v \gamma_d}{m_\chi^2 \dm^2}
\end{align}
according to eq.~\eqref{eq:defb}, such that $\langle \sigma_\text{ann} v \rangle = 6b/x$ in the non-resonant regime. By fixing $\gamma_d = 10^{-3}$, $n_d = 1$, and scanning over $\gamma_v$ and $m_\chi$, in fig.~\ref{fig:L=1results}, we show the resulting constraints in the $b-m_\chi$ parameter plane for different values of the mass splitting $\dm \in \{10^{-2}, 10^{-3}, 10^{-4}, 10^{-6}, 10^{-8}\}$ (dashed, different colors), as well as for fixed values of $\Tkd \in \{1, 10^{-2}, 10^{-4} \}\,\mathrm{MeV}$ (different panels), which have been obtained by running \acropolis with \texttt{ResonanceModel}. For comparison, we also indicate the bounds that are obtained by setting  $\langle \sigma_\text{ann} v \rangle = 6b/x$ everywhere (the ``vanilla" scenario), which we obtain by running \texttt{AnnihilationModel} instead (solid, black). 

Compared to the results obtained for $s$-wave annihilations, we again find that the constraints are different only for a finite range of masses, which becomes bigger for smaller values of $\dm$. However, in the case of $p$-wave annihilations, there does not exist a value $\dm^\text{crit}$, beyond which the constraints universally start to weaken compared to the ones with $\langle \sigma_\text{ann} v \rangle = 6b/x$. This is because, while the position of the peak remains the same, the corresponding annihilation cross-section is given by $\langle \sigma_\text{ann} v \rangle = \mathcal{O}(1) b\sqrt{\dm} /\gamma_d$, which is larger than $6b/x \sim b\dm$ for $\sqrt{\dm} \gamma_d < \mathcal{O}(1)$, i.e. for all relevant parts of parameter space. However, for some values of $\dm$, there still exist certain values of $m_\chi$, for which the bounds get \textit{weaker} compared to the vanilla scenario. This happens when $T_{\text{peak}}$ is pushed deep into the region relevant for photodisintegration. In this case, we again have $\langle \sigma_\text{ann} v \rangle \propto 1/T^{3/2}$ at $T \sim T_\text{max}$ (see above), meaning that $\langle \sigma_\text{ann} v \rangle = 6b/x$ can be larger than the resonantly-enhanced cross-section around $T_\text{max}$. Nevertheless, this effect only occurs for rather small values of $\dm \lesssim 10^{-7}$ and $\Tkd \lesssim 100\,\mathrm{eV}$ (also cf.~fig.~\ref{fig:delta}).

Additionally, for $p$-wave annihilations the resonance effect is much more pronounced, i.e.~for identical values of $\gamma_d$, $\Tkd$, and $\dm$, the enhancement relative to the vanilla scenario is orders of magnitude stronger than in the case of $s$-wave annihilations. This is because, when comparing the ratio between the resonantly-enhanced cross-section and the one in the vanilla scenario at the peak, we find for $s$-wave annihilations $\langle \sigma_\text{ann} v \rangle/a = \mathcal{O}(1)\sqrt{\dm}/\gamma_d$, while for $p$-wave annihilations we obtain $\langle \sigma_\text{ann} v \rangle /(6b/x) = \mathcal{O}(1)/(\sqrt{\dm} \gamma_d$). Consequently, in the latter case the ratio is larger by a factor $1/\dm \gg 1$, meaning that the cross-section at the peak is significantly more enhanced, which directly translates to more stringent limits.

\begin{figure}[t]
\centering
\includegraphics[width=0.495\linewidth]{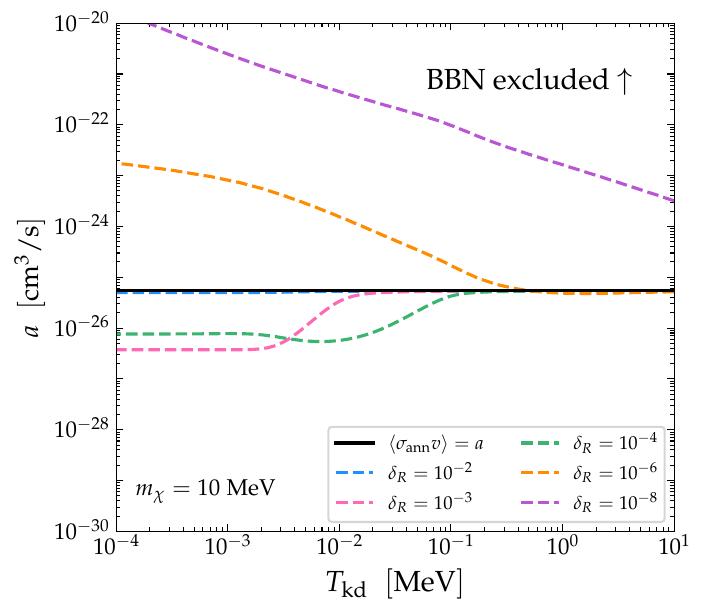}    \includegraphics[width=0.495\linewidth]{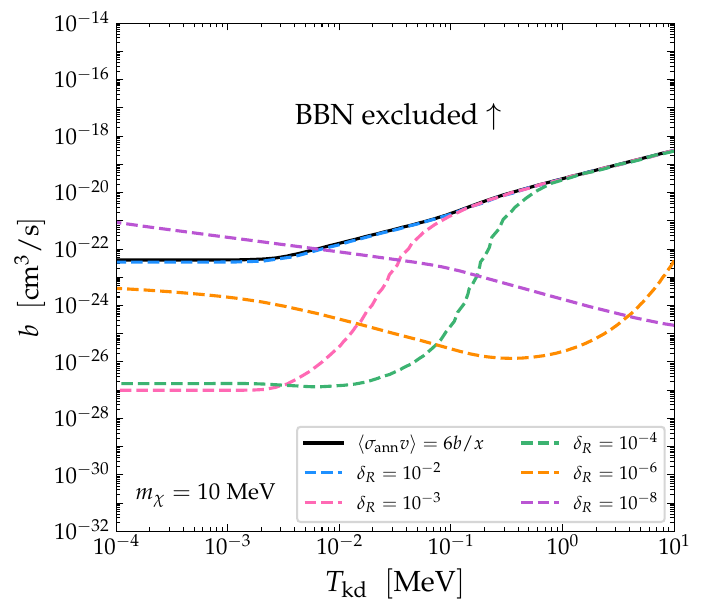}    \includegraphics[width=0.495\linewidth]{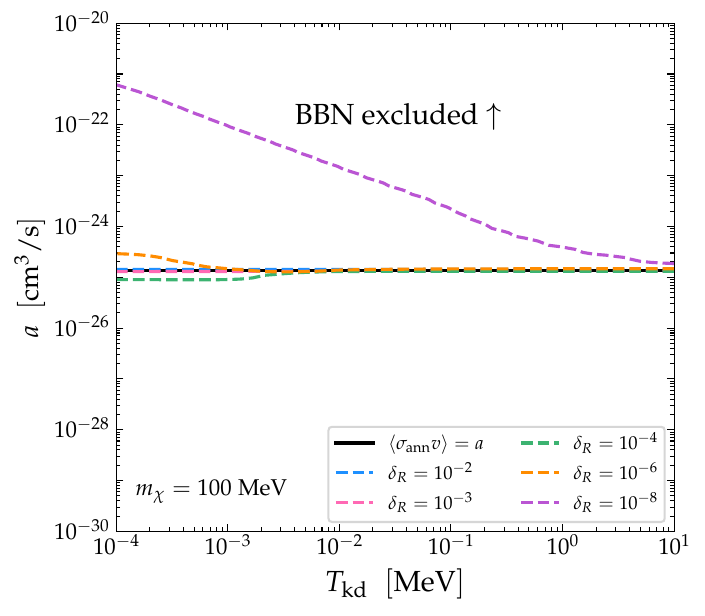}
\includegraphics[width=0.495\linewidth]{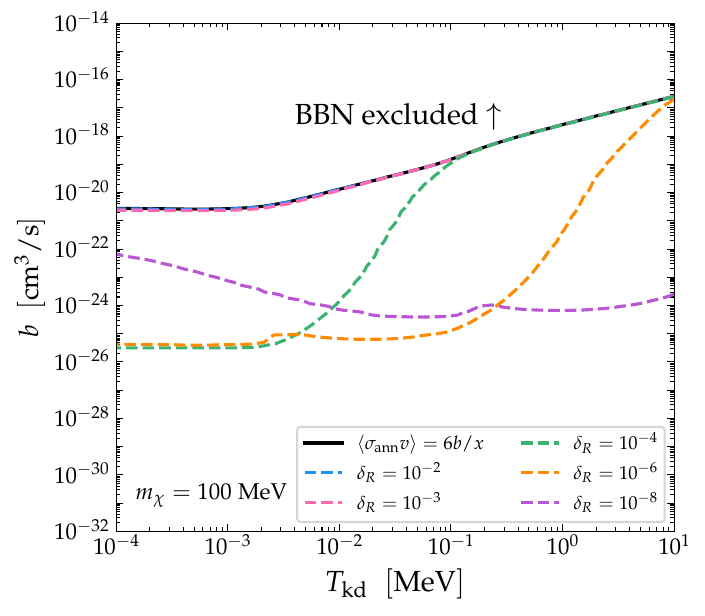}
\caption{BBN constraints from photodisintegration at 95\% C.L. for $s$-wave (left) and $p$-wave annihilations (right) as a function of the kinetic decoupling temperature $\Tkd$ for $m_\chi = 10\,\mathrm{MeV}$ (top) and $m_\chi = 100\,\mathrm{MeV}$ (bottom), as well as different values of $\dm \in \{10^{-2}, 10^{-3}, 10^{-4}, 10^{-6}, 10^{-8}\}$ (dashed, different colors). In addition, we also show the bounds on non-resonant annihilations of NR DM (solid, black) for comparison.}
\label{fig:running_Tkd}
\end{figure}

Moreover, $p$-wave constraints are also subject to a larger dependence on $\Tkd$, since $\langle \sigma_\text{ann} v \rangle \propto T_\chi$, which depends on $\Tkd$ via eq.~\eqref{eq:def_Tkin}. Specifically, the bounds strengthen significantly if the decoupling temperature is lowered from $\Tkd = 1\,\mathrm{MeV}$ to $\Tkd = 10\,\mathrm{keV}$, while not changing much afterwards. This sharp transition can be understood by means of fig.~\ref{fig:running_Tkd}, which shows the resulting constraints on $a$ and $b$ as a function of $\Tkd$ for the two masses $m_\chi = 10\,\mathrm{MeV}$ (top) and $m_\chi = 100\,\mathrm{MeV}$ (bottom), as well as for $s$-wave (left) and $p$-wave annihilations (right).\footnote{Note that for $n_d = 0$, the variation of the limits with $\Tkd$ is much less pronounced by comparison.} In general, $\langle \sigma_\text{ann} v \rangle \propto 1/x \propto T_\chi(T)$, and consequently $\langle \sigma_\text{ann} v \rangle \propto T$ ($\langle \sigma_\text{ann} v \rangle \propto T^2$) before (after) kinetic decoupling. For $T < \Tkd$, the annihilation cross-section thus falls off faster, which weakens the constraints, meaning that it is generally favourable to delay kinetic decoupling as much as possible. As the temperature of kinetic decoupling is pushed below $T_\text{max}$ at $\Tkd \sim 10^{-2}\,\mathrm{MeV}$, the cross-section stays $\propto T$ for a larger range of relevant temperatures, thus improving the limits. However, this improvement only lasts until $\Tkd \lesssim 10^{-3}$, beyond which point the bounds become independent of $\Tkd$. This is because (as discussed above), photodisintegration is most sensitive to temperatures close to $T_\text{max}$. 
Consequently, for $\Tkd \ll T_\text{max}$ the source term $\propto n_\chi^2 \langle \sigma_\text{ann} v \rangle$ is already negligible once $T < \Tkd$. This behaviour is independent of the resonance peak and can also be observed for scenarios with $\langle \sigma_\text{ann} v \rangle = 6b/x$.

\begin{figure}[t]
    \centering
    \includegraphics[width=0.65\linewidth]{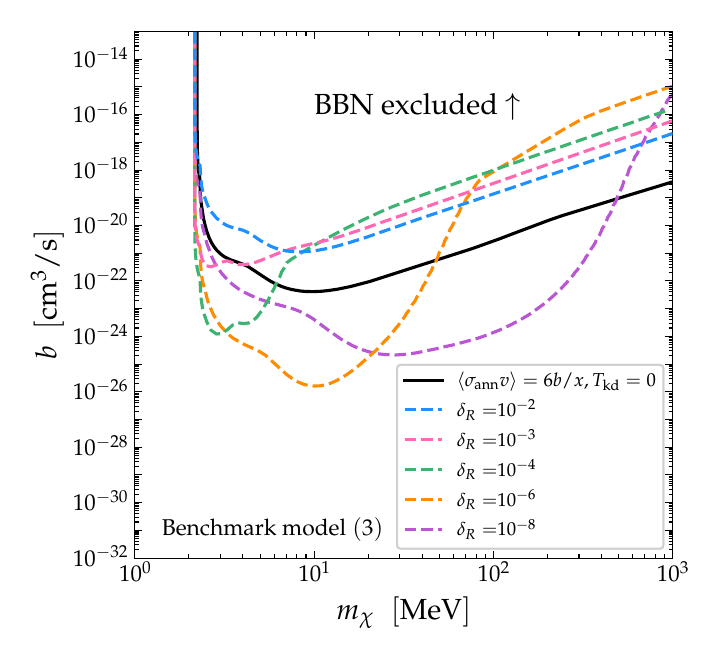}
    \caption{BBN constraints from photodisintegration at 95\% C.L. on resonant DM annihilations for $\gamma_d = 10^{-3}$, $n_d = 1$, different values of $\dm \in \{10^{-2}, 10^{-3}, 10^{-4}\}$ (dashed, different colors), and dynamically calculated values of $\Tkd$ according to eq.~\eqref{eq:def_Tkin} for the benchmark model $(3)$ of tab.~\ref{tab:models}. For comparison, we also show the (most stringent) bounds for non-resonant $p$-wave annihilations of NR DM (solid, black).}
    \label{fig:L=1results_dyn}
\end{figure}

In addition to fixed values of $\Tkd$, in fig.~\ref{fig:L=1results_dyn}, we further show the resulting constraints for dynamically calculated values of $\Tkd$ according to eq.~\eqref{eq:def_Tkin} corresponding to benchmark model $(3)$ from tab.~\ref{tab:models}, which have been obtained by running \acropolis with \texttt{BenchmarkModel3}. For comparison, we also show the (most stringent) bounds for vanilla $p$-wave annihilations (solid, black), which have been obtained by running \texttt{AnnihilationModel} with $\Tkd = 0$. Compared to the $s$-wave benchmark counterparts, the bounds from resonantly-enhanced $p$-wave annihilations differ more drastically from the vanilla scenario, which is specifically true in the low mass region, i.e.~for $m_\chi \lesssim \mathcal{O}(10)\,\mathrm{MeV}$ if $\dm \lesssim 10^{-3}$. The reason for this is twofold: On the one hand, as already described above, the resonance contribution to the annihilation cross-section is smaller in the case of $s$-wave annihilations. On the other hand, $p$-wave annihilations are generally less constrained than $s$-wave annihilations, which implies that larger values of $\gamma_v$ are still allowed. In turn, the dynamically determined values of $\Tkd$ along the bound are generally lower for $n_d = 1$ compared to $n_d = 0$ (typically $\Tkd \sim 10-100\,\mathrm{keV}$ along the bound for $p$-wave annihilations vs.~$\Tkd \sim 1 - 10\,\mathrm{MeV}$ for $s$-wave annihilations), which additionally boosts the resonance effect in line with the discussion above. Overall, we therefore conclude that, in the case of $p$-wave annihilations, it is indeed possible to strengthen the BBN bounds from photodisintegration by resonantly enhancing the annihilation cross-section, even for realistic benchmark scenarios.

\section{Conclusions}
\label{sec:conclusion}
Resonant annihilations provide an interesting avenue for boosting the DM annihilation cross-sections. However, within such scenarios, residual annihilations are usually still efficient at late times, thus injecting large amounts of electromagnetic material into the SM heat bath. The injected particles, in turn, can afterwards participate in photodisintegration reactions, thus potentially destroying some of the elements that have previously been created during BBN. Consequently, comparing the predicted abundances of light elements in such scenarios with the ones inferred from observations therefore provides a handle on the strength of the DM annihilations.

While constraints from photodisintegration have previously already been calculated for $s$-wave and $p$-annihilations in the absence of resonance effects, in this work, we derive for the first time the corresponding constraints for the case of resonantly-enhanced DM annihilations. To this end, we have implemented and made available (\url{https://github.com/hep-mh/acropolis}) a new model called \texttt{ResonanceModel} within \acropolis. This model has been implemented in a rather model-independent way, with only minimal assumptions about the DS (cf.~sec.~\ref{sec:model}). However, for concreteness, we have also implemented three different benchmark models in order to calculate constraints for more concrete scenarios. Using this new version of \acropolis, it is possible to reproduce all the results presented in this work, as well as to determine the corresponding constraints for any other combination of parameters (cf.~sec.~\ref{sec:implementation}).

By using \texttt{ResonanceModel}, we have further calculated the resulting constraints on $s$-wave and $p$-wave annihilations for \textit{(1)} fixed and \textit{(2)} dynamically calculated values of the kinetic decoupling temperature $\Tkd$ (cf.~sec.~\ref{sec:results}). In the case of $s$-wave annihilations, we find that while the constraints can be quite different from the vanilla ones for certain combinations of $\Tkd$ and $\dm$ (cf.~fig.~\ref{fig:L=0results}), this is not true for the two benchmark scenarios presented in this work (cf.~fig.~\ref{fig:L=0results2}). In fact, when correctly accounting for the kinetic decoupling temperature, we find that the constraints remain very similar to the vanilla ones. In this case, resonantly-enhanced annihilations therefore do not commonly lead to more stringent constraints. For $p$-wave annihilations, however, the constraint can be boosted for both fixed values of $\Tkd$ (cf.~fig.~\ref{fig:L=1results}), as well as for dynamically calculated values of $\Tkd$ (cf.~fig.~\ref{fig:L=1results_dyn}). It is therefore important to consider these constraints when discussing the viability of scenarios with resonantly-enhanced DM annihilations. This can be achieved by means of the new model implemented in \acropolis.

\section*{Acknowledgments}
The authors thank Marieke Postma and Torsten Bringmann for very helpful discussions. This work was funded by an NWO-klein2 grant (OCENW.KLEIN.427). The work of M.H. is further supported by the F.R.S./FNRS.

\appendix
\section{Limits for different values of $\gamma_d$}
\label{app:gammad}
\begin{figure}[t]
\centering
\includegraphics[width=0.495\linewidth]{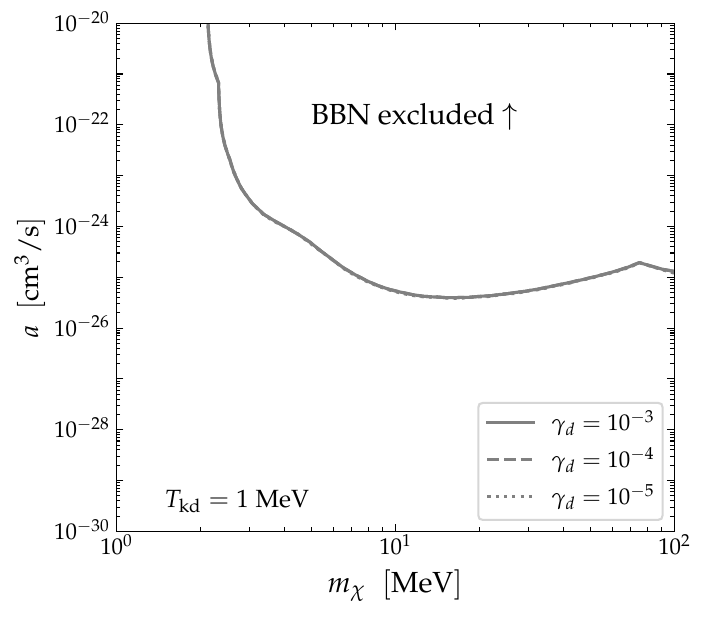}    \includegraphics[width=0.495\linewidth]{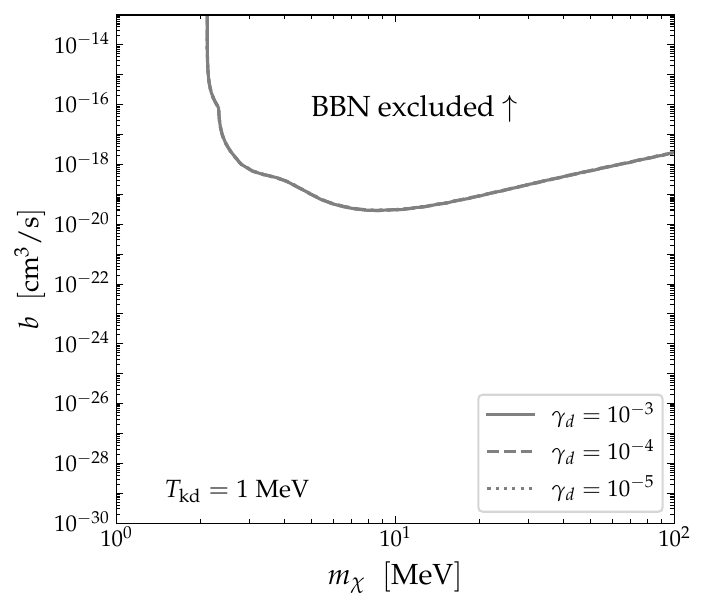}    \includegraphics[width=0.495\linewidth]{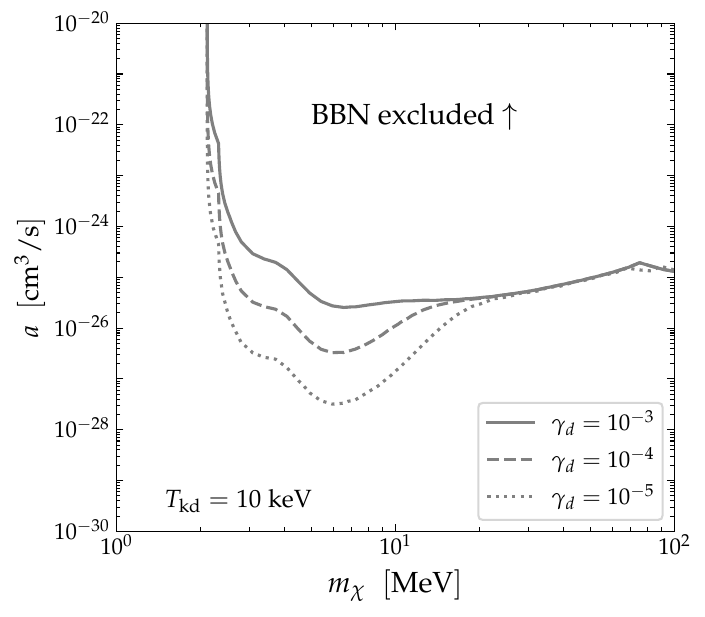}
\includegraphics[width=0.495\linewidth]{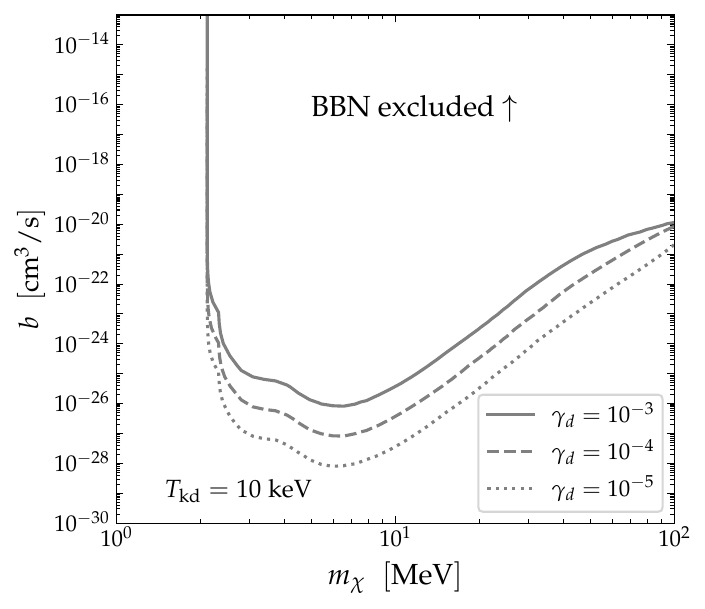}    \includegraphics[width=0.495\linewidth]{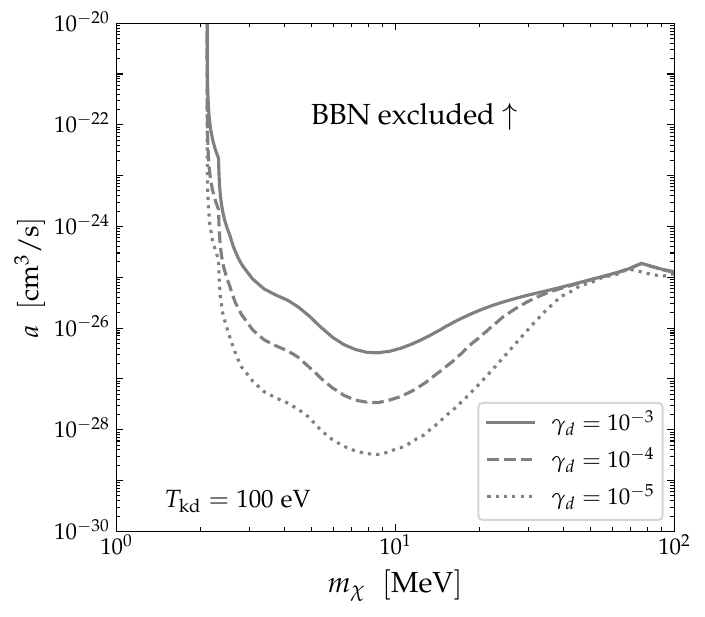}
\includegraphics[width=0.495\linewidth]{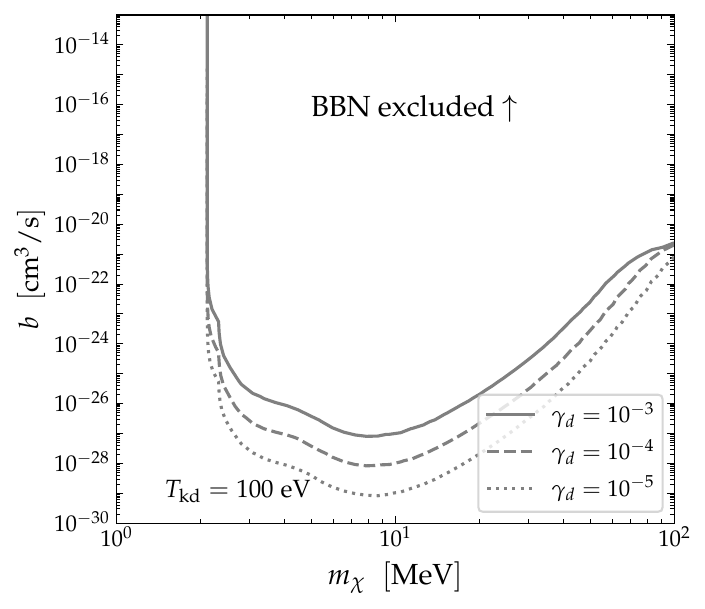}
\caption{BBN constraints from photodisintegration at 95\% C.L. for $s$-wave (left) or $p$-wave (right) annihilations for different decoupling temperatures (different panels) and different choices of $\gamma_d$ (different linestyles). Here, we explicitly set $\dm = 10^{-3}$.}
\label{fig:gd}
\end{figure}

In this paper, we have presented plots for the explicit choice $\gamma_d = 10^{-3}$. In principle, the off-resonance contribution to the annihilation cross-section is sensitive only to the combination $\gamma_d \gamma_v$. If the constraints are dominated by this contribution, it is thus possible to infer analogous limits for a different value of $\gamma_d$ by rescaling $\gamma_v$ while keeping the product of both couplings constant. However, in the limit in which one of the two couplings is much smaller than the other -- which is typically expected for physically motivated DM models --, the resonance contribution is instead sensitive to only the smallest of both couplings, and a simple rescaling is not possible if this is the dominant contribution. In order to show the effect on the limits in this case, in \cref{fig:gd} we show the resulting bound for $s$-wave (left) and $p$-wave (right) annihilations for different values of $\gamma_d \in \{ 10^{-3}, 10^{-4}, 10^{-5}\}$ (different linestyles) as well as for different values of $\Tkd \in \{ 1, 10^{-2}, 10^{-4}\}\,\mathrm{MeV}$ (different panels). If kinetic decoupling happens significantly late, the resonance effects become dominant and a simple rescaling is no longer possible. However, if the resonance-contribution is less relevant, i.e.~for sufficiently early decoupling, the constraints instead become independent of the choice of $\gamma_d$.

\newpage
\bibliographystyle{jhep} 
\bibliography{resonantDM}

\end{document}